\documentclass[graybox]{svmult}

\usepackage{mathptmx}       % selects Times Roman as basic font
\usepackage{helvet}         % selects Helvetica as sans-serif font
\usepackage{courier}        % selects Courier as typewriter font
\usepackage{type1cm}        % activate if the above 3 fonts are
\usepackage{makeidx}         % allows index generation
\usepackage{graphicx}        % standard LaTeX graphics tool
\usepackage{multicol}        % used for the two-column index
\usepackage[bottom]{footmisc}% places footnotes at page bottom

\usepackage{epsfig}
\usepackage{graphics}
\usepackage{psfrag}
\usepackage{amssymb}
\newcommand{\beq}{\begin{equation}}
\newcommand{\eeq}{\end{equation}}
\newcommand{\bea}{\begin{eqnarray}}
\newcommand{\eea}{\end{eqnarray}}

\newcommand{\e}{\mbox{e}}
\renewcommand{\d}{\mbox{d}}
\newcommand{\g}{\gamma}

\newcommand{\lam}{\lambda}
\newcommand{\La}{\Lambda}
\renewcommand{\b}{\beta}
\renewcommand{\a}{\alpha}
\newcommand{\n}{\nu}
\newcommand{\m}{\mu}

\renewcommand{\th}{\theta}
\newcommand{\del}{\delta}
\newcommand{\Del}{\Delta}
\newcommand{\sg}{\sigma}
\renewcommand{\k}{\kappa}
\newcommand{\oh}{\frac{1}{2}}

\newcommand{\ra}{\rangle}
\newcommand{\la}{\langle}
\newcommand{\prt}{\partial}
\newcommand{\mi}{\!-\!}

\newcommand{\pl}{\!+\!}

\newcommand{\cD}{{\cal D}}

\newcommand{\cN}{{\cal N}}

\renewcommand{\ts}{{\tilde{s}}}
\newcommand{\tC}{{\tilde{C}}}

\newcommand{\tL}{{\tilde{\La}}}
\newcommand{\tX}{{\tilde{X}}}
\newcommand{\tY}{{\tilde{Y}}}

\newcommand{\tc}{{\tilde{c}}}
\newcommand{\tV}{{\tilde{V}}}

\newcommand{\tk}{{\tilde{k}}}

\newcommand{\hG}{{\hat{G}}}

\newcommand{\hP}{{\hat{P}}}

\newcommand{\hPk}{\hP^{kin}}
\newcommand{\hPp}{\hP^{pot}}

\newcommand{\bN}{{\bar{N}}}

\newcommand{\SL}{\sqrt{\La}}

\newcommand{\SLT}{\sqrt{\La}T}
\newcommand{\R}{{\rm I\!R}}

\newcommand{\ointz}{\oint \frac{dz}{2\pi i \, z}\;}
\newcommand{\ointy}{\oint \frac{dy}{2\pi i \, y}\;}
\newcommand{\ointx}{\oint \frac{dx}{2\pi i \, x}\;}

\newcommand{\rf}[1]{(\ref{#1})}

\begin{document}

\title*{CDT---an Entropic Theory of Quantum Gravity}
%\titlerunning{Short Title} for an abbreviated version of
% your contribution title if the original one is too long
\author{J. Ambj\o rn,  A. G\"{o}rlich, J. Jurkiewicz and R. Loll}
% Use \authorrunning{Short Title} for an abbreviated version of
% your contribution title if the original one is too long
\institute{J. Ambj\o rn \at The Niels Bohr Institute, 
Blegdamsvej 17, DK-2100 Copenhagen \O , Denmark , \\
\email{ambjorn@nbi.dk}
\and A. G\"{o}rlich \at Institute of Physics, Jagellonian University,
Reymonta 4, PL 30-059 Krakow, Poland, \\
\email{atg@th.if.uj.edu.pl} 
\and J. Jurkiewicz \at Institute of Physics, Jagellonian University,
Reymonta 4, PL 30-059 Krakow, Poland,\\ 
\email{jurkiewicz@th.if.uj.edu.pl}
\and R. Loll \at Institute for Theoretical Physics, Utrecht University,
Leuvenlaan 4, NL-3584 CE Utrecht, The Netherlands, 
 \email{r.loll@uu.nl}}

\maketitle

\abstract{In these lectures we describe how a theory 
of quantum gravity may be constructed in terms of a lattice
formulation based on so-called causal 
dynamical triangulations (CDT). 
We discuss how the continuum limit can be obtained and how 
to define and measure diffeomorphism-invariant correlators.
In four dimensions, which has our main interest, the lattice
theory has an infrared limit which can be identified with 
de Sitter spacetime. We explain why this infrared property of 
the quantum spacetime is nontrivial and due to 
``entropic" effects encoded in the nonperturbative path integral measure.
This makes the appearance of the de Sitter universe an example of 
true emergence of classicality from microscopic quantum laws.
We also discuss nontrivial aspects of the UV behaviour, and
show how to investigate quantum fluctuations around
the emergent background geometry. Finally, we consider the connection
to the asymptotic safety scenario, and derive from it 
a new, conjectured scaling relation in CDT quantum gravity.}

\section{Introduction}\label{intro}

How to reconcile 
the classical theory of general relativity with quantum mechanics
remains an unsolved problem.
Flat Minkowskian spacetime 
seems  an excellent local approximation to spacetime down to 
the smallest distances we can probe in the laboratory. At least
the Standard Model of elementary particles, which relies heavily
on the Minkowskian spacetime structure, works almost too well,
leaving us presently with little clue as to what should replace it
at shorter spacetime distances, i.e.\ higher energies.   
If one na\"ively tries to quantize the theory of general relativity 
by making a perturbative expansion around this flat background one is 
faced with the fact that the corresponding field theory is not renormalizable.
The mass dimension of the gravitational coupling constant is $-2$ in 
units where $c= \hbar = 1$. To deal with this problem one can try to 
go beyond conventional quantum field theory. 
One such attempt is string theory. However, until now it has added 
little to our understanding of why to a very good approximation we
live in a 3+1 dimensional classical world governed by Einstein's
equations with a positive cosmological constant, 
around which there presumably are small quantum fluctuations.
Loop quantum gravity is another attempt to quantize gravity, which 
introduces new ways of treating gravity at the Planck scale, but it 
has problems with recovering classical gravity in the infrared
limit. Here we will describe a much more mundane approach using 
only standard quantum field theory. 
In a sum-over-histories approach we will attempt to 
define a nonperturbative quantum field theory which has as
its infrared limit ordinary classical general relativity 
and at the same time has a nontrivial 
ultra\-vio\-let limit. From this point of view it is close in 
spirit to the renormalization group approach, whose application to 
gravity with the hope of establishing its {\it asymptotic 
safety} was first advocated 
long ago by Weinberg \cite{weinberg},
and more recently substantiated by several groups of researchers 
\cite{reuteretc}.

The approach reported here is nontrivial for two reasons 
which combine to make 
it genuinely nonperturbative. First of all, as just stated,
a theory of quantum gravity is not perturbatively renormalizable,
and thus, whatever field theory one invents, it must in some 
sense be nonperturbative. Here we want to use a lattice to
provide an ultraviolet regularization of the quantum field theory.
The lattice regularization of quantum field theories has been 
very successful, but it is usually implemented in flat, Euclidean spacetime,
where the Osterwalder-Schrader axioms ensure us that this is unproblematic.
One knows how to get from a quantum field theory formulated in 
Euclidean spacetime to a quantum 
field theory formulated in Minkowskian spacetime.
However, little is known about the analogous issue once we move from flat 
to curved spacetimes and 
even more, to a situation where spacetime itself becomes the object of 
quantization\footnote{An extension of the Osterwald-Schrader axioms to
certain diffeomorphism-invariant theories was given in \cite{ammt}.}.
That the situation
is nontrivial even at the most elementary level is seen by considering 
the Einstein-Hilbert action from which we can derive the classical equations 
of general relativity. It is formally straightforward to rotate 
this action from Minkowskian to Euclidean signature. However, one then has
to face the fact that the 
Euclidean action is unbounded below, the unboundedness
caused by the ``wrong sign'' of the conformal mode, corresponding
to overall, local rescalings of the metric tensor.
This is a potential obstacle for constructing the quantum theory:
when trying to sum over all geometries,
with weights given by the exponential of minus the action, there are 
geometries with arbitrarily large negative action which may
render the sum over paths ill defined and the Euclidean theory nonsensical.
Consequently, the UV lattice regularization of the path integral has to 
be such that it also regularizes the infinities which can arise due to the 
conformal factor. Even if such a regularization exists (and it does, as we 
shall see), how can one ever  expect to obtain something finite
in the continuum limit where the regularization is supposed to be removed? 

The only solution from a continuum point of view is that the correct path 
integral measure in the Euclidean sector  suppresses the unbounded 
conformal factor. In the lattice approach the measure factor is of an
``entropic'' nature: it reflects how many configurations (microscopic, geometrical 
realizations) there are corresponding to a given value of the action. 
This entropic factor will enter as an integral part of the bare effective
lattice action. (We will illustrate this below in toy examples where 
everything can be calculated analytically.) The ``entropy part" of the 
effective action will be independent of its ``bare coupling constant part". 
Usually, the possibility of obtaining a continuum limit
of a lattice theory is linked to the existence of a critical point 
(more generally, a critical surface) in coupling constant space. This  
will also be the case here. The entropy part of the effective action plays a
crucial role in determining the critical value of the bare coupling constants, 
and the continuum
quantum field theory will then emerge at that critical point. However, 
in such cases there may be no ``obvious'' continuum theory one can 
read off from the lattice effective action since we might not know
the precise form of the entropy part of the effective action.
This highlights the truly nonperturbative nature of the continuum theory.  

This review article\footnote{For less technical accounts of this approach 
to quantum gravity, see \cite{desitter,causality}.} 
is organized as follows: first we describe how a two-dimensional
toy model of quantum gravity can be solved explicitly,
illustrating some of the points made above. Then we describe the
four-dimensional theory, the numerical results obtained by Monte Carlo
simulations and how to connect the lattice formalism to the renormalization 
group approach and to a new theory, so-called Ho\v rava-Lifshitz gravity.

\section{The CDT formalism in two dimensions}\label{cdt}

\subsection{Generalities}

The lattice formulation of {\it Euclidean} quantum gravity, i.e.\ the 
quantum theory of Euclidean geometries, has been very successful
in two dimensions. In two dimensions, gravity does not have any field-theoretic
degrees of freedom, but nevertheless two-dimensional Euclidean quantum
gravity constitutes a nontrivial example of a 
diffeomorphism-invariant quantum theory of geometries.
The lattice theory, regularized by the method of 
so-called dynamical triangulations (DT), provides a 
diffeomorphism-invariant cut-off of two-dimensional Euclidean quantum gravity.
It is thus a misconception that a lattice regularization  
will necessarily break diffeomorphism invariance.
Rather, one should view the use of DT in the path integral 
as a way to sum directly over {\it geometries}, thus avoiding
completely the issue of diffeomorphism invariance. The reason
why such an interpretation is possible is that the triangulations 
used in DT can be viewed as piecewise linear geometries without 
any specific metric assigned to them: once we know the lengths  
of the links and the gluing of the simplices, we have the complete 
information about the geometry. Using identical simplices 
the basic information about the geometry is entirely encoded
in the way the simplices are glued together, and the summation 
over geometries becomes the summation over possible abstract 
triangulations. The UV cut-off is the length $a$ of the sides
of the simplices. Using this formalism, one can formulate a
Euclidean theory of quantum gravity using as building blocks
Euclidean equilateral simplices and one obtains a 
lattice version of two-dimensional
quantum gravity. It can be solved analytically for finite $a$
and agrees with a continuum quantization 
of two-dimensional {\it Euclidean} 
gravity (quantum Liouville theory) in the limit $a \to 0$.

However, in spacetime dimension larger than two this Euclidean lattice 
approach does 
not seem to have the desired continuum limit. This apparent failure was a key
motivation for introducing a modified approach based on so-called 
causal dynamical triangulations (CDT). It realizes, in a nonperturbative
context, ideas put forward in earlier work \cite{teitelboim}, which advocated 
that in a gravitational path 
integral with the correct, Lorentzian signature of spacetime one should sum 
over causal geometries only.

\subsection{The combinatorial solution in two dimensions}

Let us describe the explicit solution of  CDT in two dimensions, that is,
one space and one time dimension \cite{al}. 

The model is defined as follows. The topology of the underlying manifold
is taken to be $S^1\times [0,1]$, with ``space" represented by the closed 
manifold $S^1$. We consider the evolution of this space in ``time''. 
No topology change of space is allowed. 

\begin{figure}
\centerline{\scalebox{0.6}{\rotatebox{0}{\includegraphics{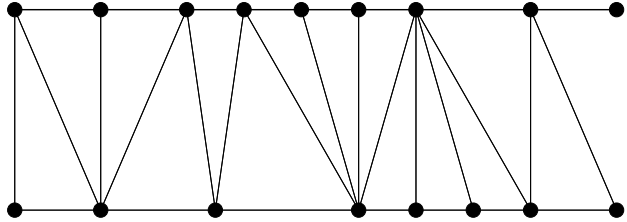}}}}
\caption[fig1xxx]{The propagation of a spatial slice from time step $t$ to 
step $t+1$ in two-dimensional causal triangulations.
The ends of the strip should be joined to form a band
with topology $S^1 \times [0,1]$.}
\label{fig1xxx}
\end{figure}

The geometry of each spatial slice is uniquely characterized by 
the length assigned to it. In the discretized version, the length $L$ 
will be quantized in units of a lattice spacing $a$, i.e.\ 
$L= l\cdot a$ where $l$ is an integer. A slice will thus be 
defined by $l$ vertices and $l$ links connecting them. 
To obtain a 2d geometry, we will evolve this spatial loop in 
discrete steps. This leads to a preferred notion of (discrete) ``time'' 
$t$, where each loop represents a slice of constant $t$.
The propagation from time-slice 
$t$ to time-slice $t+1$ is governed by the following rule: each vertex
$i$ at time $t$ is connected to $k_i$ vertices at time $t+1$, $k_i \geq 1$,
by links which are assigned a ``time-like" 
squared edge length $-a^2$. The $k_i$ vertices, $k_i > 1$, 
at time-slice $t+1$ will be connected by $k_i-1$ consecutive 
space-like links, thus forming $k_i -1$ triangles. 
Finally the right boundary vertex
in the set of $k_i$ vertices will be identified with the left boundary 
vertex of the set of $k_{i+1}$ vertices. In this way we get a total of 
$\sum_{i=1}^l (k_i-1)$ vertices (and also links) at time-slice $t+1$ and 
the two spatial slices are connected by $\sum_{i=1}^l k_i
\equiv l_{t}+l_{t+1}$ triangles,
see Fig.\ \ref{fig1xxx}. 

The elementary building blocks of a geometry are therefore triangles
with one space- and two time-like edges. We define them to be flat
in the interior. A consistent way of assigning interior angles to
such Minkowskian triangles is described in \cite{sorkin}. The
angle between two time-like edges is $\gamma_{tt}=-\arccos \frac{3}{2}$,
and between a space- and a time-like edge $\gamma_{st}=
\frac{\pi}{2}+\frac{1}{2} \arccos \frac{3}{2}$, summing up to
$\gamma_{tt}+2\gamma_{st}=\pi$. The sum over all angles around a
vertex with $j$ incoming and $k$ outgoing time-like edges (by
definition $j,k\geq 1$) is given by $2\pi+(4-j-k)\arccos\frac{3}{2}$.
The regular triangulation of flat Minkowski space
corresponds to $j=k=2$ at all vertices. The volume of a single
triangle is given by $\frac{\sqrt{5}}{4}a^2$.

One may view these geometries as a subclass of all possible
triangulations that allow for the introduction of a causal
structure. Namely, if we think of all time-like links as being
future-directed, a vertex $v'$ lies in the future of a
vertex $v$ iff there is an oriented sequence of time-like
links leading from $v$ to $v'$. Two arbitrary vertices may or may not
be causally related in this way. 

In quantum gravity one sums over all geometries connecting, 
say, two spatial boundaries of length $L_1$ and $L_2$, with the weight 
of each geometry $g$ given by 
\beq\label{1}
\e^{i S[g]}, ~~~~~S[g] = \La_0 \int \!\!\sqrt{-g}~~~(\mbox{in 2d}),
\eeq
where $\La_0$ is the bare cosmological constant. If in 
the discretized model we have a piecewise linear geometry made 
from $n$ triangles the corresponding action will be 
\beq\label{disact}
S = \La_0 \frac{\sqrt{5}a^2}{4}\;n = \lam n,~~~~
\lam \equiv  \La_0 \frac{\sqrt{5}a^2}{4}.
\eeq
In our discretized model the boundaries will be characterized by 
integers $l_1$ and $l_2$, the number of vertices or links at the two
boundaries. The path integral amplitude for the ``propagation'' from 
geometry $l_1$ to $l_2$ will be the sum over all interpolating 
surfaces of the 
kind described above, with a weight given by the discretized version of 
\rf{1}. Let us call the corresponding amplitude $G^{(1)}_\lam(l_1,l_2)$.
Thus we have
\bea
G_\lam^{(1)}(l_1,l_2) &=& \sum_{t=1}^{\infty} 
G_\lam^{(1)}(l_1,l_2;t),\label{3}\\
G_\lam^{(1)}(l_1,l_2;t) &=& 
\sum_{l=1}^\infty G_\lam^{(1)}(l_1,l;1)\;l\;G_\lam^{(1)}(l,l_2,t-1),\label{4}\\
G_\lam^{(1)}(l_1,l_2;1) &=& \frac{1}{l_1}\sum_{\{k_1,\dots,k_{l_1}\}} 
\e^{i \lam  \sum_{i=1}^{l_1} k_i}, \label{5}
\eea
where $\lam$ denotes the {\em bare}  
dimensionless lattice cosmological constant\footnote{One obtains 
the renormalized (continuum) cosmological constant $\La$  by 
an additive renormalization, see below.} defined in \rf{disact}, 
and where $t$ denotes the total number of time-slices 
connecting $l_1$ and $l_2$. 

From a combinatorial point of view it is convenient to mark a 
vertex on the entrance loop in order to get rid of the factors
$l$ and $1/l$ in \rf{4} and \rf{5}, that is,
\beq\label{6}
G_\lam (l_1,l_2;t) \equiv l_1 G_\lam^{(1)}(l_1,l_2;t).
\eeq
(The unmarking of a point may be thought of as 
the factoring out by (discrete) spatial diffeomorphisms).
Note that $G_\lam(l_1,l_2;1)$ plays the role of a 
transfer matrix, satisfying
\bea
G_\lam(l_1,l_2,t_1+t_2) 
&=& \sum_{l} G_\lam(l_1,l;t_1)\; G_\lam(l,l_2;t_2)\label{7}\\
G_\lam(l_1,l_2;t+1) &=& \sum_{l} G_{\lam}(l_1,l;1)\;G_\lam(l,l_2;t).\label{8}
\eea
Knowing $G_\lam(l_1,l_2;1)$ allows us to find $G_\lam(l_1,l_2;t)$
by iterating \rf{8} $t$ times. This program is conveniently 
carried out by introducing the generating function for the numbers
$G_\lam(l_1,l_2;t)$,
\beq\label{9}
G_\lam(x,y;t)\equiv \sum_{k,l} x^k\,y^l \;G_\lam(k,l;t),
\eeq
which we can use to rewrite \rf{7} as 
\beq\label{10}
G_\lam(x,y;t_1+t_2) = \ointz G_\lam(x,z^{-1};t_1) G_\lam(z,y;t_2),
\eeq
where the contour should be chosen to include the singularities 
in the complex $z$--plane of $G_\lam(x,z^{-1};t_1)$ but not those
of $G_\lam(z,y;t_2)$. 

One can either view the introduction of $G_\lam(x,y;t)$ as a purely
technical device or take $x$ and $y$ as related to boundary cosmological 
constants. Let $\lam_i$ and $\lam_f$ denote dimensionless lattice 
boundary cosmological constants, such that if the entrance boundary 
consists of  $k$ links the lattice boundary action will be $\lam_i k$,
or, introducing a dimensionful bare lattice boundary cosmological 
constant $\La_i = \lam_i/a$ and a continuum boundary length $L_i=k\, a$,
$\La_i L_i$ (and similarly $\La_0= \lam_0/a$ etc.). We now write 
\beq\label{10a}
x=\e^{i\lam_i}= \e^{i \La_i a},~~~~y=\e^{i\lam_f}=\e^{i\La_o a},
\eeq
such that $x^k= \e^{i\lam_i \,k}$ becomes 
the exponential of the boundary cosmological term,
and similarly for $y^l= \e^{i\lam_f \, l}$. 
Let us for notational convenience define 
\beq\label{11}
g=\e^{i\lam}.
\eeq
For the technical purpose of counting we view $x,y$ and $g$ as 
variables in the complex plane. In general the function 
\beq\label{11a}
G(x,y;g;t)\equiv G_\lam(x,y;t)
\eeq
will be analytic in a neighbourhood of $(x,y,g)=(0,0,0)$.  

From the definitions \rf{5} and \rf{6} it follows by standard techniques 
of generating functions that we may associate a factor $g$ with each 
triangle, a factor $x$ with each vertex on the entrance loop and 
a factor $y$ with each vertex on the exit loop, leading to
\beq\label{12}
G(x,y;g;1) =\sum_{k=0}^\infty \left( gx \sum_{l=0}^\infty
 (gy)^l \right)^k -
\sum_{k=0}^\infty (gx)^k = \frac{g^2 xy}{(1-gx)(1-gx-gy)}.
\eeq
Formula \rf{12} is simply a book-keeping device for all possible
ways of evolving from an entrance loop of any length in one step to
an exit loop of any length. The subtraction of the term $1/(1-gx)$ 
has been performed to 
exclude the degenerate cases where either the entrance or the exit 
loop is of length zero. 
  
From \rf{12} and eq.\ \rf{10}, with $t_1=1$, we obtain
\beq\label{13}
G(x,y;g;t) = \frac{gx}{1-gx}\; G(\frac{g}{1-gx},y;g;t-1).
\eeq
This equation can be iterated and the solution written as 
\beq\label{14}
G(x,y;g;t) = F_1^2(x)F_2^2(x) \cdots F_{t-1}^2(x) 
\frac{g^2 xy}{[1-gF_{t-1}(x)][1-gF_{t-1}(x)-gy]},
\eeq
where $F_t(x)$ is defined iteratively by
\beq\label{15}
F_t(x) = \frac{g}{1-gF_{t-1}(x)},~~~F_0(x)=x.
\eeq
Let $F$ denote the fixed point of this iterative equation. By standard
techniques one readily obtains
\beq\label{16}
F_t(x)= F\ \frac{1-xF +F^{2t-1}(x-F)}{1-xF +F^{2t+1}(x-F)},~~~~
F=\frac{1-\sqrt{1-4g^2}}{2g}.
\eeq
Inserting \rf{16} in eq.\ \rf{14}, we can write
\bea
G(x,y;g,t)\!\!\! &=&\!\!\!\!  \frac{ F^{2t}(1-F^2)^2\; xy}
{(A_t-B_tx)(A_t-B_t(x+y)+C_txy)}
\label{17}\\
~\!\!\!& =&\!\!\!\!
 \frac{F^{2t}(1-F^2)^2\;xy}{\Big[(1\!\!-\!xF)\!-\!F^{2t+1}(F\!\!-\!x)\Big]
\Big[(1\!\!-\!xF)(1\!\!-\!yF)\!-\!F^{2t} (F\!\!-\!x)(F\!\!-\!y)\Big]}~,
~~~\label{17a}
\eea
where the time-dependent coefficients are given by 
\beq\label{18}
A_t =1-F^{2t+2},~~~B_t=F(1-F^{2t}),~~~C_t=F^2(1-F^{2t-2}).
\eeq
The combined region of convergence to the 
expansion in powers $g^kx^ly^m$, valid for all $t$ is 
\beq\label{18a}
|g| < \oh,~~~~ |x|< 1,~~~~|y|<1.
\eeq
The asymmetry between $x$ and $y$ in the expressions \rf{17} and \rf{17a}
is due to the marking of the entrance loop. If we also mark the exit loop
we have to multiply $G_\lam(l_1,l_2;t)$ by $l_2$. We define
\beq\label{18b}
G_\lam^{(2)}(l_1,l_2;t) \equiv l_2\, G_\lam(l_1,l_2;t)= l_1l_2 
G^{(1)}_\lam (l_1,l_2;t).
\eeq
The corresponding generating function $G^{(2)}(x,y;g;t)$ is obtained from
$G(x,y;g;t)$ by acting with $y \frac{d}{dy}$,
\beq\label{18c}
G^{(2)}(x,y;g;t) = 
\frac{ F^{2t}(1-F^2)^2 \, xy}{(A_t -B_t(x+y)+C_t xy)^2}.
\eeq

We can compute $G_\lam(l_1,l_2;t)$ from $G(x,y;g;t)$ by
a (discrete) inverse Laplace transformation
\beq\label{19}
G_\lam(l_1,l_2;t) =\ointx \ointy \frac{1}{x^{l_1}}\, 
\frac{1}{y^{l_2}}\; G(x,y;g;t),
\eeq
where the contours should be chosen in the region where $G(x,y;g;t)$ is 
analytic. A more straightforward method is to rewrite the
right-hand side of \rf{17} as a power series in $x$ and $y$,
yielding
\beq\label{19aNEW!}
G_\lam (l_1,l_2;t) = 
\frac{F^{2t}(1-F^2)^2 B^{l_1\pl l_2}}{l_2\;\; A^{l_1\pl l_2\pl 2}} \;
\;\sum_{k=0}^{\min (l_1,l_2)\mi 1} 
\frac{l_1\pl l_2\mi k\mi 1}{k!(l_1\mi k\mi 1)!(l_2\mi k\mi 1)!}
\left( \mi \frac{A_tC_t}{B_t^2}\right)^k,
\eeq
which, as expected, is symmetric with respect to $l_{1}$ and $l_{2}$
after division by $l_{1}$. 

In the next section we will give explicit expressions  
for $G_\lam(l_1,l_2;t)$, $G_\lam(l_1,l_2)$ and $G_\lam(x,y)$ 
(the integral of $G_\lam(x,y;t)$ over $t$) in a certain continuum limit.

\subsection{The continuum limit}

The path integral formalism we are using here
is very similar to the one used to re\-pre\-sent the free particle as 
a sum over paths. Also there one performs a
summation over geometric objects (the paths), and the path integral itself
serves as the propagator. From the particle case it is known that the bare mass
undergoes an additive renormalization (even for the free particle), 
and that the bare propagator is subject to a wave-function renormalization
(see \cite{book1} for a review). The same is true
in two-dimensional gravity, treated in the formalism of 
dynamical triangulations \cite{book1}. The coupling constants
with positive mass dimension, i.e.\ the cosmological constant and the 
boundary cosmological constants, undergo an 
additive renormalization, while the partition function itself (i.e.\ the 
Hartle-Hawking-like wave functions) 
undergoes a multiplicative wave-function renormalization.
We therefore expect the bare coupling constants $\lam,\lam_i$ and 
$\lam_0$ to behave as 
\beq\label{20a}
\frac{\sqrt{5}}{4}\La_0\equiv \frac{\lam}{a^2} = \frac{C_{\lam}}{a^2} + \tilde{\La},~~~~
\La_i\equiv\frac{\lam_i}{a}= \frac{C_{\lam_{i}}}{a}+\tilde{X},~~~
\La_o\equiv\frac{\lam_f}{a} =\frac{C_{\lam_{f}}}{a}+\tilde{Y},
\eeq
where $\tilde{\La},\tilde{X},\tilde{Y}$ denote the renormalized 
cosmological and boundary cosmological constants and where we have 
absorbed a factor $\sqrt{5}/4$ in the definition of $\tilde{\La}$.

If we introduce the notation 
\beq\label{20c}
g_c = \e^{i C_{\lam}},~~~~x_c= \e^{i C_{\lam_{i}}},~~~~y_c=\e^{iC_{\lam_{f}}},
\eeq
for critical values of the coupling constants, 
it follows from \rf{10a} and \rf{11} that 
\beq\label{20b}
g=g_c\,\e^{ia^2\tL},~~~~x=x_c\,\e^{ia\tX},~~~~y=y_c\,\e^{ia\tY}.
\eeq 
The wave-function renormalization will appear as a multiplicative  
cut-off dependent factor in front of the bare 
``Green's function'' $G(x,y;g;t)$, 
\beq\label{20}
G_\tL (\tX,\tY;T) = \lim_{a \to 0} a^{\eta} G(x,y;g;t),
\eeq
where $T=a\, t$, and where the critical exponent $\eta$ 
should be chosen such that 
the right-hand side of eq.\ \rf{20} exists. In general this will only be 
possible for particular
choices of $g_c$, $x_c$ and $y_c$ in \rf{20}. 

The basic relation \rf{7} can survive the limit \rf{20} only 
if $\eta=1$, since we have assumed that the boundary lengths 
$L_1$ and $L_2$ have canonical dimensions and satisfy $L_i = a\, l_i$. 

From eqs.\ \rf{17} and \rf{18} it is clear that we can only 
obtain a nontrivial continuum limit if $|F| \to 1$.
This leads to a one-parameter family of possible choices
\beq\label{23}
g_c= \frac{1}{2\cos \a}~~~~\mbox{for}~~~F=\e^{i\a},~~~ \a\in \R,
\eeq
for critical values of $g$. 
It follows from \rf{11} that most values of $g_c$ correspond 
to a complex {\em bare} cosmological constant $\lam$. However, the 
renormalized cosmological constant $\tilde{\La}$ in 
\rf{20a} (depending on how we approach $g_c$ in the complex plane)
could in principle still be real.  

\begin{figure}
\centerline{\scalebox{0.6}{\rotatebox{0}{\includegraphics{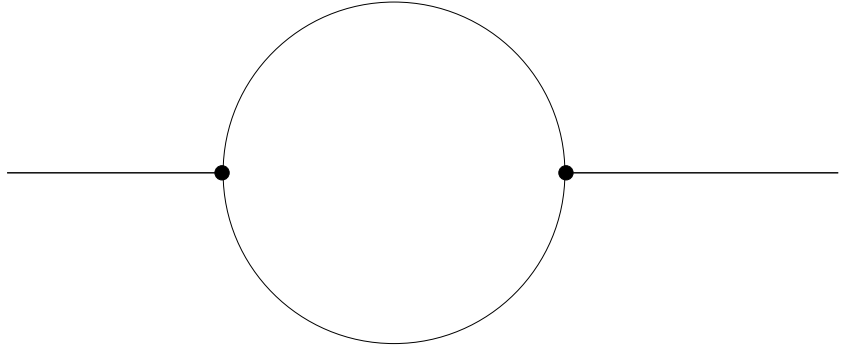}}}}
\caption[fig2xxx]{The circle of convergence in 
the complex $g$ plane (radius 1/2),
and the critical lines, ending in $g=\pm 1/2$.}
\label{fig2xxx}
\end{figure}

A closer analysis reveals that only at $g_c=\pm 1/2$, 
corresponding to $\a=0,\pi $, is there 
any possibility of obtaining an interesting continuum limit.
Note that these two values are the only ones  
which can be reached from a region of 
convergence of $G(x,y;g;t)$ (see Fig.\ \ref{fig2xxx}).
Note also that requiring the bare $\lambda$ to lie inside 
the region of convergence when $g \to g_c$ leads to a restriction 
Im$\, \tL > 0$ on the {\em renormalized} cosmological constant $\tL$,
since $|g|<\frac{1}{2}\Rightarrow \mbox{Im}\ \lambda >\ln 2$.

Without loss of generality, we will consider the critical value
$g_c=1/2$. It corresponds to 
a purely imaginary bare cosmological constant
$\lam_{c}:=C_{\lam}/a^{2} = -i \ln 2/a^2$.
If we want to approach this point from the region in the 
complex $g$-plane where $G(x,y;g;t)$  
converges it is natural to choose the renormalized coupling $\tL$
imaginary as well, $\tL = i\La$, i.e.
\beq\label{25a}
\lam = i\ \frac{\ln \oh}{a^2} +i \La.
\eeq
One obtains a well-defined scaling limit (corresponding to
$\La\in \R$) by letting $\lam\to \lam_{c}$ along the imaginary axis.
The Lorentzian form for the continuum 
propagator is obtained by an analytic continuation $\La\to -i\La$
in the {\it renormalized} coupling of the resulting Euclidean 
expressions. 

At this stage it may seem that we are surreptitiously reverting
to a fully Euclidean model. We could of course equivalently 
have conducted
the entire discussion up to this point in the ``Euclidean sector'',
by omitting the factor of $-i$ in the exponential \rf{1} of the
action, choosing $\lam$ positive real and taking all edge lengths 
equal to 1. However, from a purely Euclidean point of view there
would not have been any reason for restricting the state sum to a subclass
of geometries admitting a causal structure. The associated 
preferred notion of a discrete time allows us to define
an ``analytic continuation in time'' (we will discuss this in more detail
later for higher-dimensional gravity). Because of the simple
form of the action in two dimensions, the rotation 
\beq\label{25b}
\int  dx\ dt \sqrt{-g_{lor}} \to i\int  dx \ dt_{eu} \sqrt{g_{eu}}
\eeq
to Euclidean metrics in our model is equivalent to the analytic continuation
of the cosmological constant $\La$. What is special about the above
situation  is that we perform the analytic continuation configuration  by 
configuration, i.e.\ geometry by geometry. That is 
possible because of the particular set of causal geometries we 
have chosen to include in the regularized path integral. Moreover, as 
will be clear later, it is a feature which extends to higher dimensions too:
each piecewise linear geometry with Lorentzian signature 
we use in the path integral has an analytic 
continuation to a Euclidean piecewise linear geometry {\it and} one
has a relation like \rf{25b}  for the Einstein-Hilbert actions of the
two geometries (see later for details).

From \rf{17} or \rf{17a} it 
follows that we can only get macroscopic loops in the limit 
$a \to 0$ if we simultaneously take $x,y \to 1$. (For $g_c=-1/2$, one
needs to take $x,y \to -1$. The continuum expressions one obtains
are identical to those for $g_c=1/2$.) Again 
the critical points correspond to purely imaginary 
bare boundary cosmological coupling constants. We will 
allow for such imaginary couplings and thus approach the 
critical point $\lam_i= \lam_f=0$ from the region of convergence of 
$G(x,y;g;t)$, i.e.\ via real, positive $X,Y$ where  
\beq\label{25c}
\lam_i = i X a,~~~~\lam_f=i Y a.
\eeq
Again $X$ and $Y$ have an obvious interpretation as positive boundary
cosmological constants in a Euclidean theory, which may be
analytically continued to imaginary values to reach the Lorentzian
sector.

Summarizing, we have 
\beq\label{25}
g=\oh \e^{-\La a^2} \to \oh (1-\oh \La a^2),~~~(\mbox{i.e.}~~F=1-a\sqrt{\La})
\eeq
as well as 
\beq\label{25d}
x=\e^{-Xa} \to 1-aX,~~~~~~y=\e^{-aY} \to 1-aY,
\eeq
where the arrows $\to$ in \rf{25} and \rf{25d} should be viewed 
as analytic coupling constant redefinitions of $\La,X$ and $Y$,
which we have performed to get rid of factors of 1/2 etc. in the formulas
below.
With the definitions \rf{25} and \rf{25d} it is straightforward 
to perform the continuum limit of $G(x,y;g,t)$ as $(x,y,g) \to 
(x_c,y_c,g_c)=(1,1,1/2)$, yielding
\bea
G_\La(X,Y;T) &=& \frac{4\La\ \e^{-2\SLT}}{(\SL+X)+\e^{-2\SLT}(\SL-X)}
\nonumber\\
&&\times \, \frac{1}{(\SL+X)(\SL+Y)-\e^{-2\SLT}(\SL-X)(\SL-Y)}.
\label{26}
\eea
For $T \to \infty$ one finds 
\beq\label{27}
G_\La(X,Y;T) \buildrel{T\rightarrow
\infty}\over\longrightarrow \frac{4\La \;\e^{-2\SLT}}{(X+\SL)^2(Y+\SL)}.
\eeq

From $G_\La(X,Y;T)$
we can finally calculate $G_\La(L_1,L_2;T)$, the continuum 
amplitude for propagation from a loop of length $L_1$, 
with one marked point, at time-slice $T=0$ to a loop of length $L_2$ 
at time-slice $T$, by an inverse Laplace transformation,
\beq\label{22}
G_\La(L_1,L_2;T) = \int_{-i\infty}^{i\infty} d X \int_{-i\infty}^{i\infty} d Y
\; \e^{X L_1}\;\e^{Y L_2}\; G_\La(X,Y;T).
\eeq
This transformation can be viewed as the limit of \rf{19} for 
$a \to 0$. The continuum version of \rf{10} thus reads
\beq\label{22a}
G_\La(X,Y;T_1+T_2) = \int_{-i\infty}^{i\infty}d Z \; 
G_\La(X,-Z;T_1) \, \,G_\La(Z,Y;T_2),
\eeq
where it is understood that the complex contour of integration 
should be chosen to the left of 
singularities of $G_\La(X,-Z;T_{1})$, but to the right of those of 
$G_\La(Z,Y,T_{2})$. 
By an inverse Laplace transformation we get in the limit
$T\rightarrow\infty$
\beq\label{27a}
G_\La(L_1,L_2;T) \buildrel{T\rightarrow
\infty}\over\longrightarrow  4 L_1 \e^{-\SL (L_1+L_2)} \;\e^{-2\SLT},
\eeq
where the origin of the factor $L_1$ is the marking of a 
point in the entrance loop. For $T\to 0$ we obtain
\beq\label{28}
G_\La(X,Y;T) \buildrel{T\rightarrow
0}\over\longrightarrow \frac{1}{X+Y},
\eeq
in agreement with the expectation that the inverse Laplace transform
should behave like
\beq\label{29}
G_\La(L_1,L_2;T) \buildrel{T\rightarrow
0}\over\longrightarrow \delta(L_1-L_2).
\eeq
The general expression for $G_\La(L_1,L_2;T)$ can be computed
as the inverse Laplace transform 
of formula \rf{26}, yielding
\beq\label{30}
G_\La(L_1,L_2;T) = \frac{\e^{-[\coth \SLT] \SL(L_1+L_2)}}{\sinh \SLT}
\; \frac{\sqrt{\La L_1 L_2}}{L_2}\; \; 
I_1\left(\frac{2\sqrt{\La L_1 L_2}}{\sinh \SLT}\right), 
\eeq
where $I_1(x)$ is a modified Bessel function of the first kind.
The asymmetry between $L_1$ and $L_2$ arises because the entrance loop 
has a marked point, whereas the exit loop has not. The amplitude with 
both loops marked is obtained by multiplying with $L_2$, while the 
amplitude with no marked loops is obtained after dividing 
\rf{30} by $L_1$. The highly nontrivial 
expression \rf{30} agrees
with the loop propagator obtained from a bona-fide continuum calculation
in proper-time gauge of pure 2d gravity by Nakayama \cite{nakayama}.

The important point we want to emphasize here is that the additive 
renormalization of the cosmological constant is an entropic effect 
when calculated after rotation to Euclidean signature. 
In fact, we can write the propagator \rf{11a} as 
\beq\label{ny1}
G(x,y,g;t)= \sum_{k,l,n}x^ky^lg^n \sum_{T(k,l,n)} \frac{1}{C(T)},
\eeq
where the summation is over all {\it causal} triangulations $T(k,l,n)$ 
(as defined above and rotated to  Euclidean signature), consisting of 
$n$ triangles and with the two boundaries made of $k$ and $l$ links.
$C(T)$ is the order of the so-called automorphism group of 
graph $T$ and in our case, with a mark on one boundary, $C(T) = 1$.
The critical point is $g_c=1/2$. That can only be the case because the 
number of (causal) triangulations constructed from  $n$ triangles 
grows exponentially as $e^{n \ln 2}$. The continuum renormalized 
cosmological constant, as defined by eq.\ \rf{25}, emerges when taking the 
difference between the value of the action for a geometry made of 
$n$ triangles and the {\it entropy} of the configurations with a given 
action (which in this case is proportional to the number of triangles
$n$. More precisely, let the number of causal triangulations 
which can be constructed from 
$n$ triangles be 
\beq\label{ny2}
\cN(n) = f(n)\, \e^{\lam_c n},~~~~~\lam_c= \ln 2,
\eeq
where $f(n)$ is a prefactor growing slower than exponentially, and which 
can also depend on the boundary cosmological constants $x,y$, a dependence
we will suppress here. We can now write eq.\ \rf{ny1} as
\beq\label{ny3}
G(\lam) = \sum_n f(n)\; \e^{-(\lam - \lam_c)n}, ~~~g\equiv\e^{-\lam}.
\eeq
Introducing the notation $A=n a^2$ for the continuum area (again 
disposing of a factor $\sqrt{5}/4$ for notational simplicity) we see that 
\rf{25} can be written as
\beq\label{ny4}
\lam = \lam_c + \La a^2,
\eeq
introducing the renormalized cosmological constant $\La$.
Eq.\ \rf{ny3} can now be written as 
\beq\label{ny5}
G(\La) = \int_0^\infty \d A \; f(A/a^2) \;\e^{-\La A},
\eeq
with the continuum action $\La A$ and the nontrivial physics
contained in the function $f(A/a^2)$.

The two-dimensional CDT model can be generalized in a number of ways:
one can use different weights and explore the universality of the model
\cite{charlotte} and there exists a Hamiltonian formulation \cite{bergfin}.
Matter can be coupled to the model \cite{cdtmatter} and one can 
weaken the causal constraints \cite{generalizations}. In addition, one 
can relax the constraint of a bounded geometry \cite{d-branes}.

\section{Causal dynamical triangulations in four dimensions}

\subsection{The choice of triangulations}

The generalization from two spacetime dimensions to three or four
is in principle straightforward
\cite{3d,ajl4d}.
In what follows we will concentrate on the 4d case.
We consider spacetimes with the topology $[0,1]\times S^3$. In principle
we can choose any spatial topology, as long as we do not allow it to change
during time evolution. Here,
for simplicity, we will always take the topology of space to be that of a
three-sphere. 

Suppose now that we have a foliation of spacetime where ``time'' 
is taken to mean proper time. Each time-slice, with the topology of $S^3$, is 
represented by a three-dimensional triangulation. We choose as the set of 
possible triangulations of $S^3$ those which can be constructed
from gluing together tetrahedra whose links are all of length $a_s=a$,
playing the role of lattice spacing and UV cut-off.
These tetrahedra are thus building blocks for our curved $S^3$-geometries,
which we take to be piecewise linear. The curvature of such a piecewise linear
geometry is located at the links. A number of tetrahedra will share a link.
Each tetrahedon has a dihedral angle associated with that link, and the sum
of dihedral angles of the tetrahedra sharing the link would add up to $2\pi$ 
if space was flat around that link. If the dihedral angles add up to something
different it signals that the piecewise linear space is not flat.
In our case the tetrahedra are all identical with a dihedral angle 
$\th_d = \arccos (1/3)$, which implies there is no exact tessellation of flat 
three-dimensional space using equilateral tetrahedra. However, it is not 
important for the use we are making of the piecewise linear geometries: we use 
them in the path integral where we sum over all geometries (of a given, fixed topology). 
Thus the important question is whether the set of piecewise linear 
geometries we are using is dense in the set of geometries relevant for the 
path integral. To answer this question, we need to know the measure 
on the set of geometries.
Presently we do not even have a mathematical characterization of the
set of geometries to be used in the path integral (and the path 
integral has of course not been defined in any mathematical sense). 
Will the set of relevant geometries be like the set of paths used 
in the path integral of the particle, i.e.\ all continuous 
paths? Will the path integral over geometries include all 
``continuous'' geometries? Because the answer is presently unknown, we will
proceed with a straightforward generalization of the class of piecewise straight
paths of the particle case and see what we get.

We now connect two neighbouring 
$S^3$-triangulations $T_3(1)$ and $T_3(2)$, associated with
two consecutive discrete proper times 1 and 2, and create a four-dimensional,
piecewise linear geometry,
such that the corresponding four-dimensional 
``slab'' consists of four-simplices, has the topology of $[0,1]\times S^3$,
and has $T_3(1)$ and $T_3(2)$ as its three-dimensional boundaries.
The spatial links (and subsimplices) contained in these 
four-dimensional simplices lie in either $T_3(1)$ or $T_3(2)$, 
and the remaining links are 
time-like with proper length squared $a_t^2 = -\a a^2$, $\a >0$. 
Subsimplices which contain at least 
one time-like link we will call ``time-like". 
In discrete units, we can say that $T_3(1)$ and $T_3(2)$ are separated by a 
single step in time direction, corresponding to a time-like
distance $\sqrt{\a} a$ in the sense that each link in the slab
which connects
the two boundaries has a squared proper length $-\a a^2$. It does not imply that 
all points on the piecewise linear manifold defined by $T_3(1)$ have 
a proper distance squared $-\a a^2$ to the piecewise linear manifold defined 
by $T_3(2)$ in the piecewise Minkowskian metric of the triangulation.

\begin{figure}[t]
%\vspace{0.5cm}
\centerline{\scalebox{0.5}{\rotatebox{0}{\includegraphics{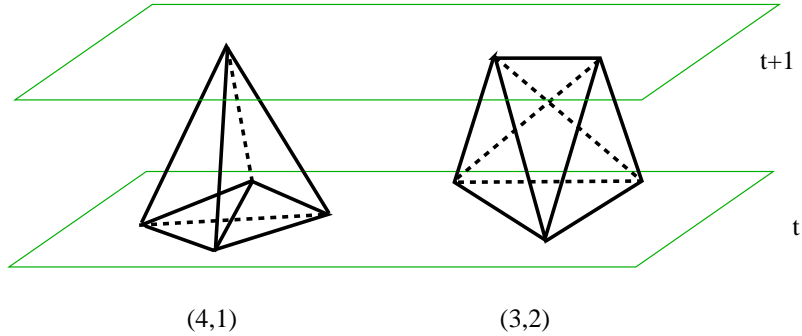}}}}
\caption{A $(4,1)$- and a $(3,2)$-simplex interpolating between 
two neighbouring spatial 
slices. The time-reversed $(1,4)$- and $(2,3)$-simplices are obtained by
turning these upside down.}
\label{connect}
\end{figure}

Thus, our slabs or ``sandwiches" are assembled from four-dimensional simplicial
building blocks of four kinds, which are labelled according to the
number of vertices they share with the two adjacent spatial slices
of constant integer proper time. A $(4,1)$-simplex has one tetrahedron (and
consequently four vertices) in common with $T_3(1)$, and only one
vertex in common with $T_3(2)$. It has four 
time-like links, connecting each of the four vertices in $T_3(1)$ 
to the vertex belonging to $T_3(2)$. In the second 
kind of four-simplex, of type $(1,4)$, the roles of $T_3(1)$ and $T_3(2)$ 
are interchanged. By contrast, a $(3,2)$-simplex has a spatial triangle (and
consequently three vertices) in common with the slice $T_3(1)$ and a spatial
link (with two vertices) in common with $T_3(2)$, together with six time-like links
connecting the two slices. The corresponding $(2,3)$-simplex is again obtained by 
interchanging $T_3(1)$ and  $T_3(2)$. The allowed simplices, up to time reversal, are
shown in Fig.\ \ref{connect}. For our purposes, we need not keep track of
the numbers of four-simplices and their time-reversed counterparts separately, and
will denote the
total number of $(4,1)$- and $(1,4)$-simplices by $N^{(4,1)}_4$ and 
similarly the total number of $(3,2)$- and $(2,3)$-simplices by $N_4^{(3,2)}$.
An allowed four-dimensional triangulation of the
slab has topology $[0,1]\times S^3$, is a simplicial manifold with boundary, 
and is constructed according to the recipe above. 
To summarize, a path in the gravitational path integral consists of a 
sequence of triangulations of $S^3$, denoted 
by $T_3(k)$, $k=0,\ldots,n$, where the space between each pair
$T_3(k)$ and $T_3(k+1)$ has been filled in by a layer of four-simplices.
In the path integral 
we sum over all possible sequences $\{T_3(k)\}$ and all possible
ways of triangulating the slabs in between $T_3(k)$ and $T_3(k+1)$.

\subsection{The choice of action}   

Piecewise linear geometries allow for a natural realization of the 
Einstein-Hilbert action, as discovered by Regge \cite{regge}. For a
piecewise linear geometry in $d$ dimensions, 
defined by a triangulation with length assignments to each link,
the curvature is concentrated on the $(d-2)$-dimensional sub-simplices. 
For example, in three dimensions the curvature is located at the links, 
as described above. In dimension four, the curvature
is concentrated at the triangles of the triangulation. 
A direct measure of the sectional curvature of the subspace perpendicular
to a given triangle is the {\it deficit angle} one can associate with it,
defined as the 
difference between the sum of ``dihedral'' angles
of the four-simplices sharing the triangle and $2\pi$. A deficit angle different from
zero signifies nonvanishing curvature.
In our case we have 
two kinds of triangles, purely space-like ones and time-like ones
(where two of the links are time-like). The local contribution to the total 
integrated curvature from such a $(d-2)$-dimensional sub-simplex is 
the volume of the sub-simplex multiplied by the deficit angle around it, 
and the integrated
discrete (scalar) curvature action is then the sum of these contributions. 
This leads to a discretized Einstein-Hilbert action of the form  
\begin{eqnarray}
S^{\rm EH}&=& \frac{1}{16\pi G}\int \d^4 x \sqrt{-g(x)} \;\Big( R(x) -
2\La \Big) ~~~~~\to
\nonumber\\
S^{\rm Regge}&=&k\;\Big(
\sum_{\stackrel{\rm space-like}{ \triangle}}{\rm Vol}(\triangle) 
\ \frac{1}{i}
\Big(2\pi -\!\!\!\!\!\!\sum_{\stackrel{\rm 4-simplices}{{\rm at}\ \triangle}}
\Theta\Big)+
\sum_{\stackrel{\rm time-like}{\triangle}}{\rm Vol}(\triangle)\  
\Big(2\pi - \!\!\!\!\!\!
\sum_{\stackrel{\rm 4-simplices}{{\rm at}\ \triangle}}
\Theta\Big)\Big)\nonumber\\
&&-\lam \;\Big(  \sum_{\stackrel{\rm (4,1)\& (1,4)-}{\rm tetrahedra}}
{\rm Vol}(4,1) +\!\!\!
\sum_{\stackrel{\rm (3,2) \& (2,3)-}{\rm tetrahedra}}{\rm Vol}(3,2)\Big),
\label{act4dis1}
\end{eqnarray}
where $R$ is the Ricci scalar curvature and $\La$ the cosmological constant.
Furthermore, we can read off that the constant $k$ is proportional to the 
inverse of the gravitational 
coupling constant $G$ and the constant $\lam$ is proportional to  $\La/G$. 

It is straightforward to calculate the volumes of the spatial and 
time-like triangles as well as the four-volumes of the $(4,1)$- and 
$(3,2)$-simplices. This enables us to express the discretized Einstein-Hilbert
action as function of the numbers $N_4^{(4,1)}$ and $N^{(3,2)}$ defined in
the previous subsection and the total number $N_0$ of vertices in the 
triangulation, leading to
\begin{eqnarray}
S^{\rm Regge} &=& (N_0 -\chi)\cdot {k \pi} \sqrt{4\alpha +1}\  
+ \mbox{\hspace{7cm}}\nonumber\\
&&
N_{4}^{(4,1)}\cdot\Biggl(k \frac{\pi}{2}\sqrt{4\alpha +1} -
\sqrt{3} k\ {\rm arcsinh} \frac{1}{2\sqrt{2}\sqrt{3\alpha +1}}\nonumber\\
&&~~~~~~~~~~~~~~ 
-\frac{3 k}{2} \sqrt{4\alpha +1} \arccos\frac{2\alpha +1}{2 (3\alpha +1)}
-\lambda \frac{\sqrt{8\alpha +3}}{96}
\Biggr)+\nonumber \\
&& N_{4}^{(3,2)}\cdot
\Biggl( k\pi \sqrt{4\alpha +1}+
\frac{\sqrt{3} k}{4}\ {\rm arcsinh}\frac{\sqrt{3}\sqrt{12 \alpha +7}}
{2 (3 \alpha+1)} - \nonumber \\
&&~~~~~~~~~~~~~~\frac{3 k}{4} \sqrt{4\alpha +1}
\biggl( 2\arccos\frac{-1}{2\sqrt{2}\sqrt{2\alpha +1}\sqrt{3\alpha +1}} +
\arccos\frac{4\alpha +3}{4 (2\alpha +1)}\biggr)\nonumber \\ 
&&~~~~~~~~~~~~~~-\lambda \frac{\sqrt{12\alpha +7}}{96}\Biggr).
\label{4dloract}
\end{eqnarray}
For details of the derivation of this formula from \rf{act4dis1}
we refer to \cite{ajl4d}. The quantity $\chi$ denotes the Euler characteristic of the 
four-dimensional spacetime and appears because we have been using 
the Euler relation $N_0-N_1+N_2-N_3+N_4= \chi$ along with other so-called
Dehn-Somerville 
relations to express the numbers $N_1$, $N_2$ and $N_3$ in terms of
$N_4$ and $N_0$, where $N_i$ counts the number of $i$-dimensional
(sub-)simplices of a given triangulation. The constant $\alpha$ comes from
allowing for a finite scaling $a_t^2 = -\a a_s^2$ between the length assignment  
of the proper length of space- and time-like links. 

The Lorentzian action (\ref{4dloract}) is obviously real for $\a >0$. 
We now want to 
study its rotation to Euclidean signature. This is naturally implemented
by performing 
an analytic continuation in $\a$ from positive to negative $\a$.
In this way, the squared proper lengths of all links become positive and we have 
a piecewise linear geometry of Euclidean signature, where
the links connecting two vertices from neighbouring time slices
have length $a_t^2 = |\a| a_s^2$. Of course, we also want 
the Euclideanized four-simplices to be nondegenerate. This requires 
$-\a > 7/12$, the
value below which a $(3,2)$-simplex becomes degenerate. Performing the 
analytic continuation in the complex lower $\a$-half-plane and 
ending at a negative value smaller than -7/12 results in the following 
Euclidean action ({\it note that we have made a redefinition $\a \to -\a$,
such that now $\a > 7/12$}): 
\bea
&&\!\!\! S_E^{\rm Regge}=-k\pi \sqrt{4\alpha -1}(N_0-\chi)\nonumber\\
&&+ N_{4}^{(4,1)} \Biggl( k\sqrt{4\alpha -1} \biggl(  -\frac{\pi}{2}
-\frac{\sqrt{3}}{\sqrt{4\alpha -1}}
 \arcsin\frac{1}{2\sqrt{2}\sqrt{3\alpha -1}} \nonumber\\
&& \;\;\;\;\;\;\;\;\;\;\;\;\;\;\; +\frac{3}{2}
\arccos\frac{2\alpha -1}{6 \alpha -2}\biggr)+
\lambda \frac{\sqrt{8\alpha -3}}{96} \Biggr)\nonumber\\
&&+N_{4}^{(3,2)} \Biggl( k \sqrt{4\alpha -1} \biggl( -\pi
+\frac{\sqrt{3}}{4\sqrt{4\alpha -1}}\arccos
\frac{6\alpha -5}{6 \alpha -2}
+\frac{3}{4} \arccos\frac{4\alpha -3}{8\alpha -4}
\nonumber\\
&&\;\;\;\;\;\;\;\;\;\;\;\;\;\;\; +\frac{3}{2}
\arccos\frac{1}{2\sqrt{2}\sqrt{2\alpha -1}\sqrt{3\alpha -1}}\biggr)
+\lambda \frac{\sqrt{12\alpha -7}}{96} \Biggr). 
\label{act4dis}
\eea
This action is precisely the Regge action for a piecewise linear geometry,
constructed from Euclidean building blocks where $a_t^2 = |\a| a_s^2$.
By analytic continuation in the complex lower-half $\a$-plane we have 
therefore arrived at the usual formula
\beq\label{ny10}
i S_{Minkowski} \to -S_{Euclidean}
\eeq
from quantum field theory,
where $S_{Euclidean}$ expressed in terms of (continuum) Euclidean geometry is
\beq\label{ny10a}
S_{E}^{\rm EH} = -\frac{1}{16\pi G} \int \d^4 \xi\, \sqrt{g(\xi)}
\; \Big( R(\xi)-2\La \Big).
\eeq 
In quantum field theory on a flat, Minkowskian background
such a ``Wick rotation" does not usually refer to a 
specific configuration in the path integral. A configuration will typically
not even be differentiable and therefore not have any analytic 
continuation. Rather, it will refer to the functional form of the action
$S[\phi]$ and how it 
changes by formally replacing $ it_M$ by $t_E$. This reflects the fact 
that the Feynman propagator, i.e.\ the result of a functional integration
over the field(s) $\phi$, has specific analytic properties which 
allow for a rotation in $t$ to the corresponding Euclidean propagator.
In our framework the situation is different, in that each individual 
piecewise linear geometry can be rotated to 
a corresponding Euclidean geometry and its associated action 
transforms in accordance with \rf{ny10}.
   
\begin{figure}[t]
%\centerline{\scalebox{1.0}{\rotatebox{0}{\includegraphics{alfa-delta_correct.ep%s}}}}
\centering
\includegraphics[width=1.0\textwidth]{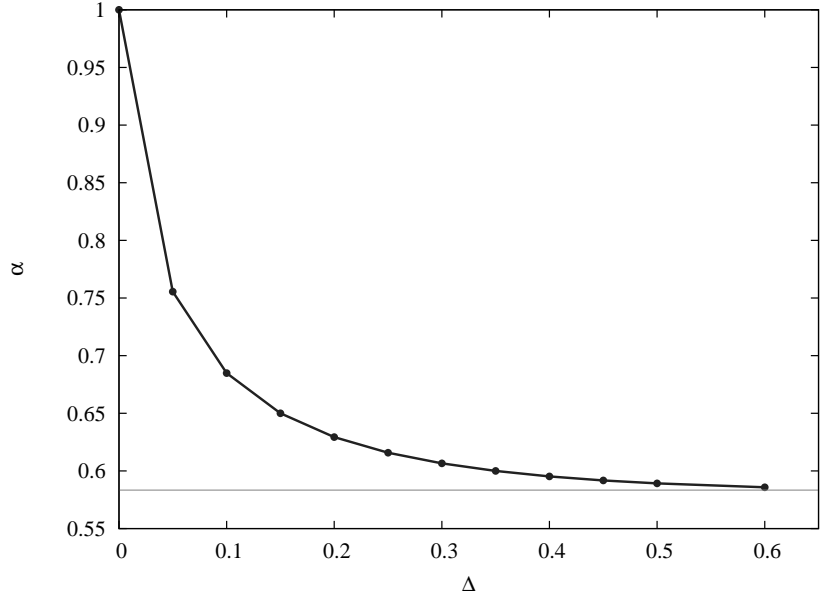}
\caption{\label{alfa} Plot of the asymmetry factor $\alpha$, 
defined as $a_t^2 = \a a_s^2$, where $a_t$ and $a_s$ are the lengths
of time-like and space-like links, plotted as a function of $\Del$.
The horizontal line is $\a=7/12$, the lowest allowed value of $\a$, where
the (3,2)-simplices collapse in the time direction.  
}
\end{figure}

For what follows it will be convenient to simplify this action and 
write it as 
\beq
 S_E^{\rm Regge}= 
-(\kappa_0+6\Delta) N_0+\kappa_4 (N_{4}^{(4,1)}+N_{4}^{(3,2)})+
\Delta (2 N_{4}^{(4,1)}+N_{4}^{(3,2)}),
\label{actshort}
\eeq 
where $\k_0=k \pi \sqrt{4\a-1}$, $\k_4$ is a linear combination of 
$k$ and $\lam$ with coefficients depending of $\a$, and $\Del$ is, 
for fixed $\k_0$ and $\k_4$, a function of $\a$, chosen such that 
for $\a=1$ (i.e.\ when $a_t=a_s$) we have $\Del=0$. 
For the values of $\k_0$ and $\k_4$ which will be of interest 
for us any $\Del >0$ corresponds to $a_t < a_s$. In Fig. \ref{alfa}
we show $\a$ as function of $\Del$ for $\k_0= 2.2$ and $\k_4$ chosen
critical (a concept to be discussed shortly). Furthermore, we have dropped
the term involving $\chi$. It plays no role in the dynamics
since we are not changing the topology.

\subsection{The entropic theory of gravity and its phase diagram}

Having chosen the cut-off and an action, we can now write down 
the path integral or partition function for the CDT version of 
quantum gravity,
\beq\label{2.1}
Z(G,\La) = \int \cD [g] \; \e^{-S_E^{\rm EH}[g]} ~~~\to~~~ 
Z(\k_0,\k_4,\Del) = 
\sum_{T} \frac{1}{C_T} \; \e^{-S_E(T)},
\eeq
where the summation is over all causal triangulations $T$ of the kind 
described above, and we have dropped the superscript ``Regge" on 
the discretized action $S_E$ given by \rf{actshort}. 
Like in the two-dimensional model, the factor $1/C_T$ is a  
symmetry factor, given by the inverse of the order $C_T$ of 
the automorphism group of the triangulation $T$. 
Note that we can write the partition function as
\beq\label{ny11}
Z(\k_0,\k_4,\Del) = \sum_{N_4,N_4^{(4,1)},N_0} \e^{-(\k_4+\Del) N_4} \;
\e^{-\Del N_4^{(4,1)}} \;\e^{(\k_0+6\Del)N_0}\!\!\!\!\!
\sum_{T(N_4,N_4^{(4,1)},N_0)}  \frac{1}{C_T} .
\eeq
Introducing 
\beq\label{ny12}
x=\e^{-(\k_4+\Del)},~~~~y=\e^{-\Del},~~~~z=\e^{(\k_0+6\Del)} ,
\eeq
we can write 
\beq\label{ny13}
\tilde Z(x,y,z) = \sum_{N_4,N_4^{(4,1)},N_0} x^{N_4}\;y^{N_4^{(4,1)}}\;z^{N_0}
\; \cN(N_4,N_4^{(4,1)},N_0),
\eeq
where $\cN(N_4,N_4^{(4,1)},N_0)$ denotes the number of CDT configurations
with $N_4$ four-simplices of which $N_4^{(4,1)}$ are of type $(4,1)$ or 
$(1,4)$, and with $N_0$ vertices, including symmetry factors.
Thus the calculation of the partition function is in principle a
combinatorial problem, just as in two dimensions where 
we could solve the problem explicitly (and formula \rf{ny13} is similar to \rf{ny1}).
This is the reason why we call the model entropic:
{\it the partition function is entirely determined by the number of geometries
in the simplest possible way, namely, by being the generating function 
for these numbers.} The counting of geometric ``microscopic" configurations
of given $(N_4,N_4^{(4,1)},N_0)$ is their entropy in the sense of statistical models. 
Unlike in two dimensions, it has until now not been possible to solve this 
counting problem analytically, which means that we will
have to rely on numerical methods. 

\begin{figure}[t]
%\vspace{0.5cm}
\centerline{\scalebox{0.6}{\rotatebox{0}{\includegraphics{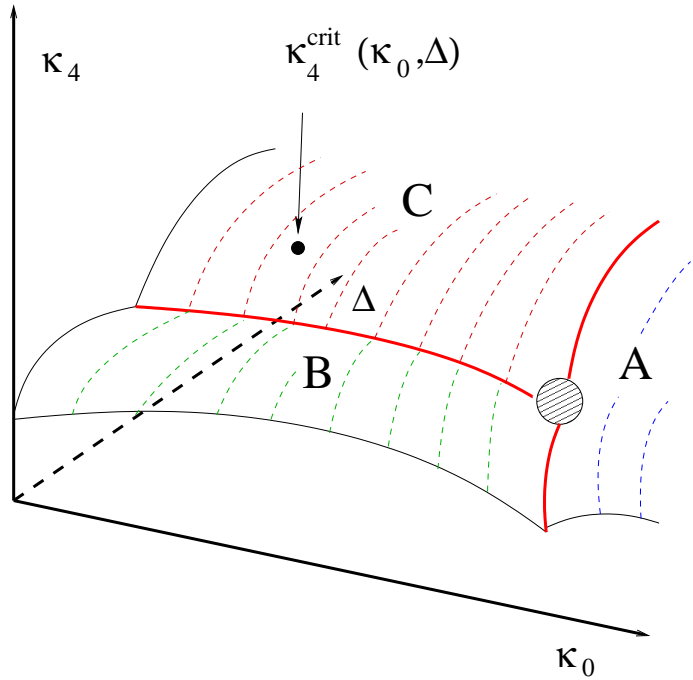}}}}
\caption{The phases A, B and C in the coupling constant space spanned by 
$(\k_0,\Del,\k_4)$. Phase C is the one where extended 
four-dimensional geometries emerge. }
\label{ph-diagram}
\end{figure}

Let us first understand better the nature of the partition function given by 
\rf{ny11}. We can write the sum as 
\beq\label{ny14}
Z(\k_0,\k_4,\Del) = \sum_{N_4} \e^{-(\k_4+\Del) N_4} \; Z_{N_4}(\k_0,\Del),
\eeq
where $Z_{N_4}(\k_0,\Del)$ is the partition function for a fixed number 
$N_4$ of four-simplices, namely,
\beq\label{ny15}
Z_{N_4}(\k_0,\Del)= \sum_{T_{N_4}}  \frac{1}{C_T} 
\e^{-\Del N_4^{(4,1)}(T_{N_4})} \;\e^{(\k_0+6\Del)N_0(T_{N_4})}.
\eeq 
One can show that $Z_{N_4}(\k_0,\Del)$ is exponentially bounded as a function
of $N_4$ \cite{expobound},
\beq\label{ny16}
Z_{N_4}(\k_0,\Del) \leq \e^{(\k_4^c+\Del)N_4}f(N_4,\k_0,\Del),
\eeq
where $f(N_4)$ grows slower than exponentially. We call $\k_4^c$ the 
{\it critical} value of $\k_4$. It is a function of $\Del$ and $\k_0$
and plays the same role as $\lam_c$ in the two-dimensional model discussed 
above: the partition function is only defined for $\k_4 > \k_4^c$ and
the ``infinite-volume'' limit, where $\la N_4\ra \to \infty$ can only 
be achieved for $\k_4 \to \k_4^c$. We are interested in sending the 
lattice spacings $a=a_s,a_t$ to zero while keeping the physical 
four-volume, which is roughly $N_4 a^4$, fixed. Thus 
we want to consider the limit $N_4 \to \infty$, and fine-tune
$\k_4$ to $\k_4^c$ for fixed $\k_0,\Del$. This fine-tuning is similar
to the fine-tuning $\lam \to \lam_c$ in the two-dimensional model. 
Like there, we expect the {\it physical} cosmological constant 
$\Lambda$ to be defined 
by the {\it approach} to the critical point according to
\beq\label{ny17}
\k_4 = \k_4^c + \frac{\La}{16\pi G} a^4,
\eeq
an equation similar to \rf{ny4}. It ensures that the term
\beq\label{ny17a}
(\k_4-\k_4^c)\;N_4 =  \frac{\La}{16\pi G} V_4,~~~~V_4=N_4a^4,
\eeq
gives rise to the standard cosmological term in the Einstein-Hilbert action.
The corresponding phase diagram is described qualitatively 
by Fig.\ \ref{ph-diagram}. The shaded surface 
is the ``critical surface'', which we want to approach from above 
(by decreasing $\k_4$). 
We put ``critical surface'' in quotation marks since more correctly 
fine-tuning to this surface corresponds to taking the limit of infinite
four-volume, which does {\it not necessarily} imply also a continuum limit.
The situation here may be different from that of the 
two-dimensional model, where
approaching $\lam_c$ automatically meant taking a continuum limit.
Two-dimensional quantum gravity is of course a very simple model with 
no propagating degrees of freedom, whereas in four-dimensional quantum gravity 
we expect to have genuinely propagating field degrees of 
freedom. Thus the situation
is more like in ordinary Euclidean lattice field theory/critical phenomena, 
where ``infinite volume'' does not
necessarily mean ``continuum limit''.  

A good example of what one might
expect is the Ising model on a finite lattice. To obtain a phase
transition for this model one has to take the lattice volume
to infinity, since there are no genuine phase transitions for finite
systems. However, just taking the lattice volume to infinity is not
sufficient to ensure critical behaviour of the Ising model. We also 
have to tune the coupling constant to its critical value, at which 
point the spin-spin correlation length diverges.
Similarly, having placed ourselves on the ``critical'' surface of 
CDT quantum gravity 
or, rather, its ``infinite-volume'' surface, 
we can discuss the various phases, indicated
in the figure as A, B and C. 
We move between these phases by 
changing the bare coupling constants $\Del$ and $\k_0$. The solid lines 
drawn on the surface are the phase transition lines between the 
(potentially) different phases, which can be of
first or higher order. In order to go from the discrete lattice 
to the continuum theory, we are often interested in a second- or 
higher-order phase transition, since such transitions are associated with 
divergent correlation lengths of the field propagators, allowing one to 
forget about the lattice spacing relative to the correlation length.
We will be looking for such transition points.

How can one imagine obtaining an interesting continuum behaviour 
as a function of $\k_0$? For the purpose of illustration, let us
assume that the subleading correction $f(N_4,k_0)$ has the form
\beq\label{ny17b}
f(N_4,\k_0) = \e^{k (\k_0) \sqrt{N_4}},
\eeq
(where later we will check numerically that such a term is indeed present, cf. eq.\ 
(\ref{n7bb})).
The partition function now has the form
\beq\label{ny17x}
Z(\k_4,\k_0) = \sum_{N_4} \e^{-(\k_4-\k_4^c)N_4 + k(\k_0) \sqrt{N_4}}.    
\eeq
For dimensional reasons we expect the classical Einstein term in the action
to scale like
\beq\label{ny17c}
\frac{1}{16 \pi G} \int \d^4 \xi \, \sqrt{g(\xi)}\;\;R(\xi) 
\;\;\;\propto \;\;\;\frac{\sqrt{V_4}}{G},
\eeq
motivating the search for a value $\k_0^c$ with
$k(\k_0^c) =0$, with the approach to this point governed by
\beq\label{ny17d}
k(\k_0)\propto \frac{a^2}{ G},~~~~{\rm i.e.}~~~~~
k(\k_0)\sqrt{N_4}\;\propto \;\frac{\sqrt{V_4}}{G}.
\eeq
With such a choice we can identify a continuum limit where 
$\la N_4\ra$, calculated from \rf{ny17x} 
(by a trivial saddle point calculation), goes to infinity while $a \to 0$,
\beq\label{ny17e}
\la N_4\ra = \frac{ \sum_{N_4} N_4 \; 
\e^{-(\k_4-\k_4^c)N_4 + k(\k_0) \sqrt{N_4}}}{\sum_{N_4} 
\e^{-(\k_4-\k_4^c)N_4 + k(\k_0) \sqrt{N_4}}} \approx 
\frac{k^2(\k_0)}{4(\k_4-\k_4^c)^2} \propto \frac{1}{\La^2 a^4}.
\eeq
Thus we find 
\beq\label{ny17f}
\la V_4\ra \propto \frac{1}{\La^2},~~~~Z(\k_4,\k_0) \approx 
\exp\Big(\frac{k^2(\k_0)}{4(\k_4-\k_4^c)}\Big) =  
\exp\Big(\frac{c}{G\La}\Big),
\eeq
as one would na\"ively expect from Einstein's equations, with the partition
function being dominated by a typical instanton contribution, for a
suitable constant $c$. 

The actual set-up for the computer simulations is slightly 
different from the theoretical framework discussed above, in that we choose 
to work with a fixed number of four-simplices $N_4$ in the 
computer simulations, rather than fine-tuning $\k_4$ to its critical 
value. We can perform computer simulations for various $N_4$
(and fixed $\k_0,\Del$) and in this way check scaling with respect
to $N_4$. This is an alternative to fine-tuning $\k_4$,
and much more convenient from a computational point of view. For large 
$N_4$ we can then check whether there are any finite-size effects or 
whether 
effectively we already are at the ``critical surface'' shown in 
Fig.\ \ref{ph-diagram}. In addition, we fix  
the total number $N_t$ of spatial slices, with proper-time labels $t_1$,
$t_2= t_1 + a_t$, up to $t_{N_t} = t_1 + (N_t\mi 1) a_t$, 
where $\Delta t\equiv a_t$ is the discrete lattice
spacing in the temporal direction\footnote{The separation between 
adjacent slices is $a_t$, in the sense that all links connecting 
two neighbouring slices have length $a_t$, as discussed above.}, 
and denote by $T = N_t a_t$ the 
total extension of the universe in proper time. 
For convenience we identify $t_{N+1}$
with $t_1$, in this way imposing the topology $S^1\times S^3$ rather
than $[0,1]\times S^3$. This choice does not affect physical results,
as will become clear below when we present the numerical results.

Finally, the computer simulations are so-called Monte Carlo simulations,
where the computer simply generates a sequence of configurations, in this 
case piecewise linear geometries, with the correct probability distribution,
as dictated by the measure and action used in the path integral. It is 
based on a local updating in terms of ``moves", which change the 
piecewise 
linear geometry in a well-defined way (see \cite{ajl4d} for a detailed 
description). These geometries are used to calculate expectation 
values of observables.

\subsection{The actual phase diagram}

\begin{figure}[t]
\center
\scalebox{0.45}{\rotatebox{-90}{\includegraphics{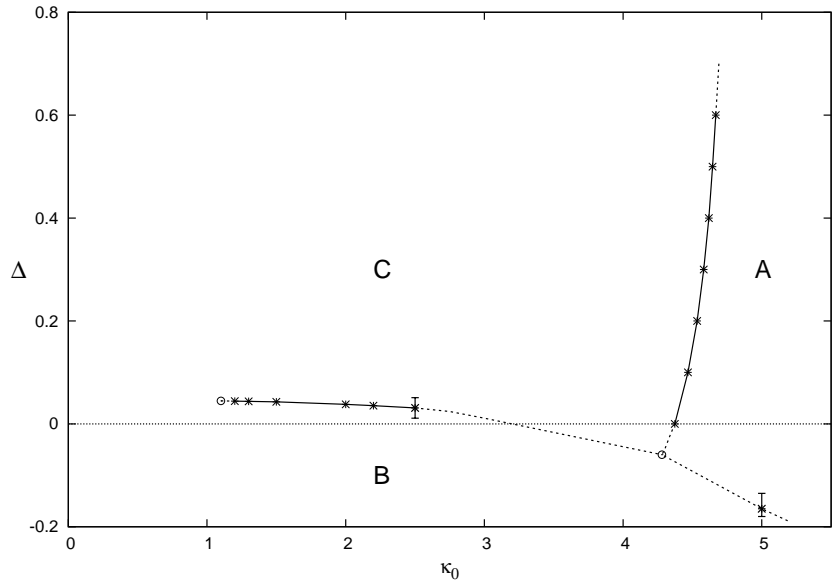}}}
\caption{The phase diagram of four-dimensional quantum gravity, defined
in terms of causal dynamical triangulations, parametrized by the inverse
bare gravitational coupling $\kappa_0$ and the asymmetry parameter $\Delta$.} 
\label{figphdiagram}
\end{figure}

Based on computer simulations with $N_4=80.000$ we have constructed
the phase diagram shown in Fig.\ \ref{figphdiagram} \cite{horava}.
The dotted lines in the figure
represent mere extra\-po\-lations,
and lie in a region of coupling constant space which is difficult to access due
to the inefficiencies of our computer algorithms. 

There are three phases, labeled A, B and C. 
In phase C, which had our main interest in \cite{emerge,blp,bigs4,agjl}, we 
observe a genuinely four-dimensional universe in the sense that 
as a function of the continuum four-volume $V_4$ (linearly related to
the number of four-simplices),
the time extent scales as $V^{1/4}_4$ and the spatial volume 
as $V_4^{3/4}$. Moving into phase A, these 
scaling relations break down. Instead,
we observe a number of small universes arranged along the time 
direction like ``pearls on a string", if somewhat uneven in size.
Individual beads along the time direction 
can grow and shrink, be created or disappear
as a function of the Monte Carlo time used in the simulations.
These small universes are connected by thin ``necks'', i.e. slices 
of constant integer time $t_n$, 
where the spatial $S^3$-universes are at or close to the smallest
three-volume permitted (consisting of five tetrahedra glued together),
thus preventing ``time" from becoming disconnected. 

By contrast, phase B is characterized 
by the ``vanishing" of the time direction, in the sense that 
only one spatial hypersurface has a three-volume appreciably larger than 
the minimal cut-off size of five just mentioned. 
One might be tempted to conclude that the resulting universe is 
three-dimensional, just lacking the time direction of the extended
universe found in phase C. However, the situation is
more involved; although we have a large three-volume collected at
a single spatial hypersurface, the corresponding spatial universe has almost 
no extension in the spatial directions. 
This follows from the fact (ascertained through direct measurement)
that it is possible to get in just a few steps 
from one tetrahedron to any other by moving along the centres of neighbouring 
tetrahedra or, alternatively,
from one vertex to any other along a chain of links. 
The Hausdorff dimension is therefore quite high, and possibly infinite. 
Let us assume for the moment that it is indeed infinite; then the universe 
in phase B has neither time nor spatial extension, and there is
no geometry in any classical sense.  

\begin{figure}[t]
%\vspace{-0.5cm}
%\begin{flushleft}
\centerline{{\scalebox{0.5}{\rotatebox{90}{\includegraphics{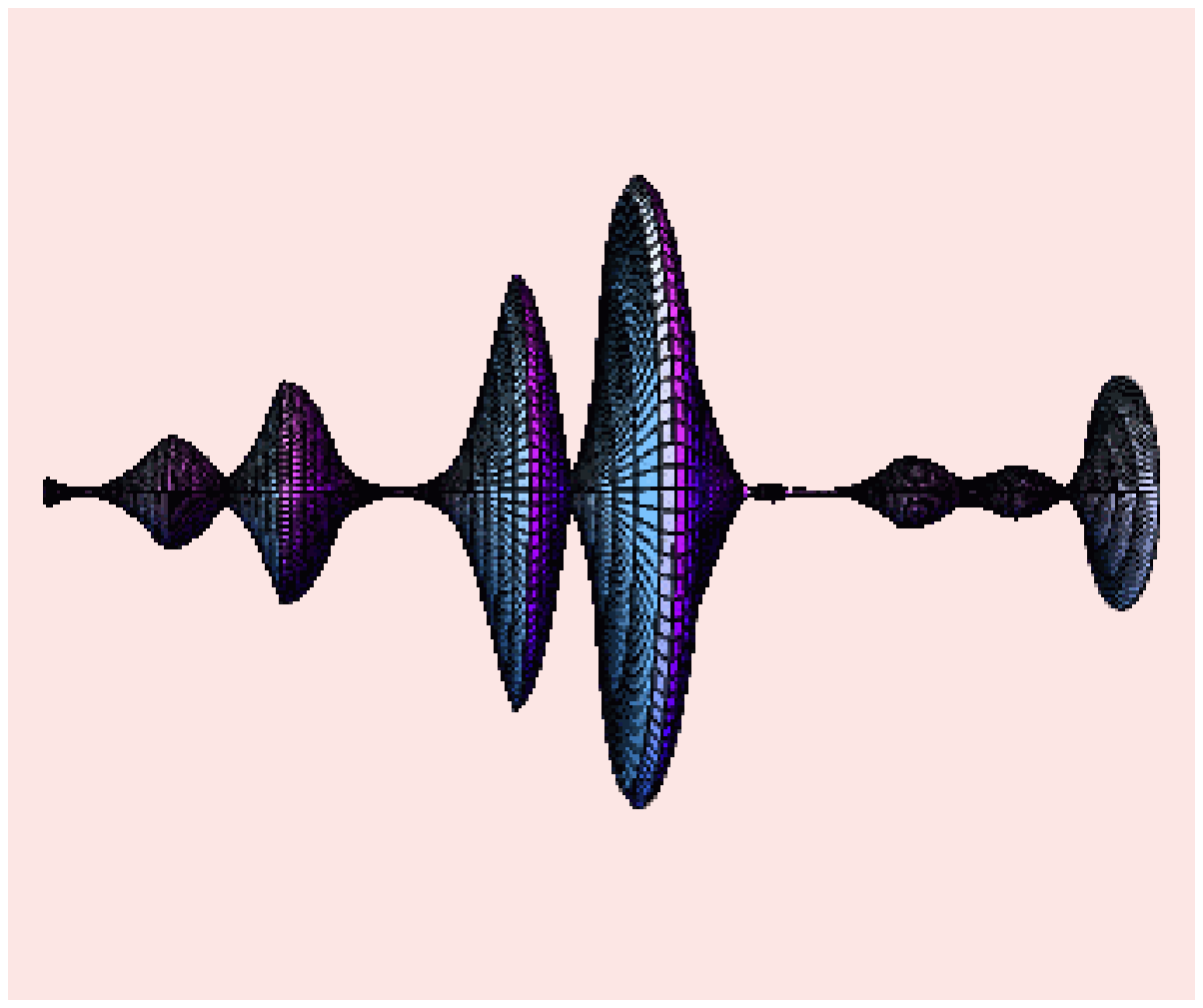}}}}
{\scalebox{0.5}{\rotatebox{90}{\includegraphics{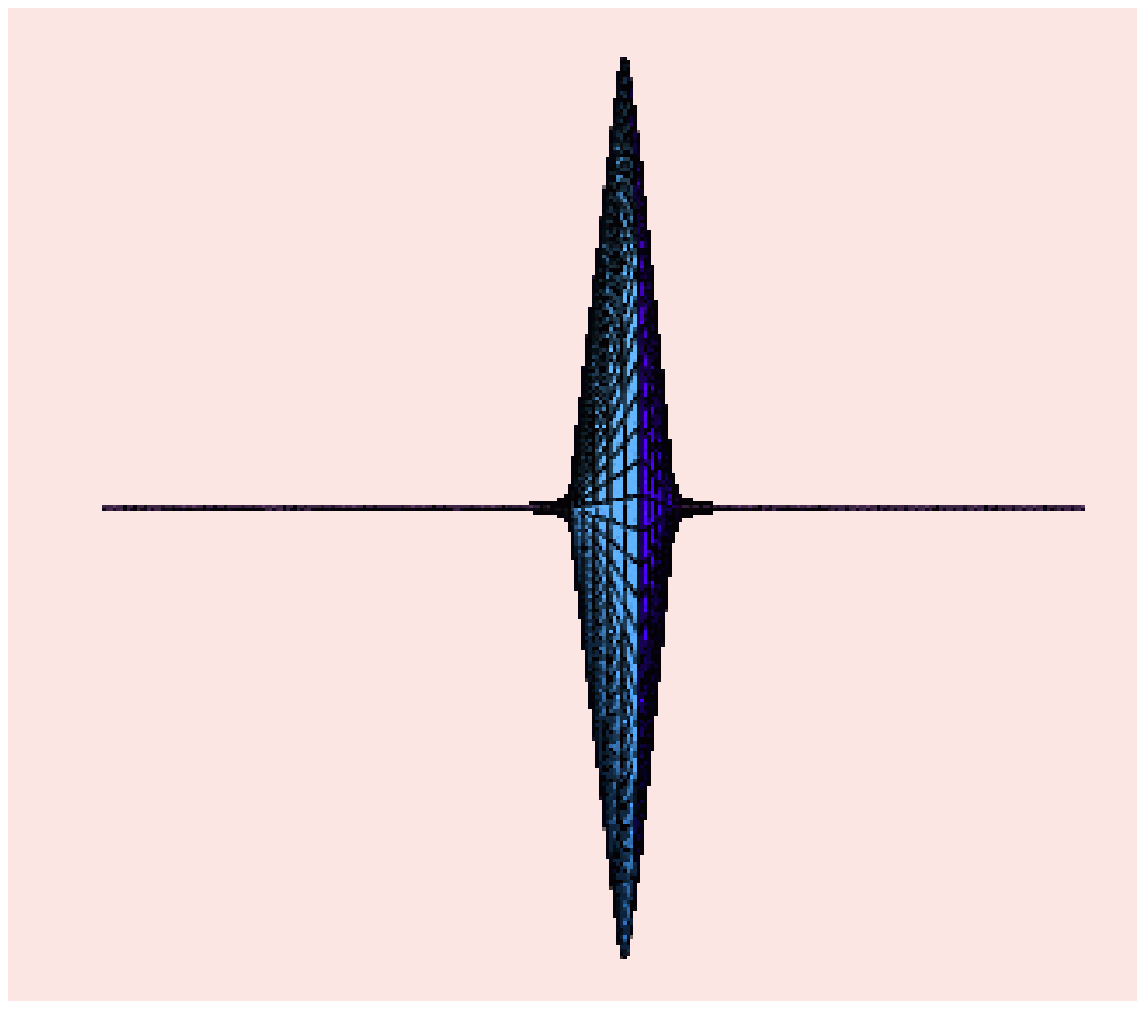}}}}}
\centerline{{\scalebox{0.712}{\rotatebox{90}{\includegraphics{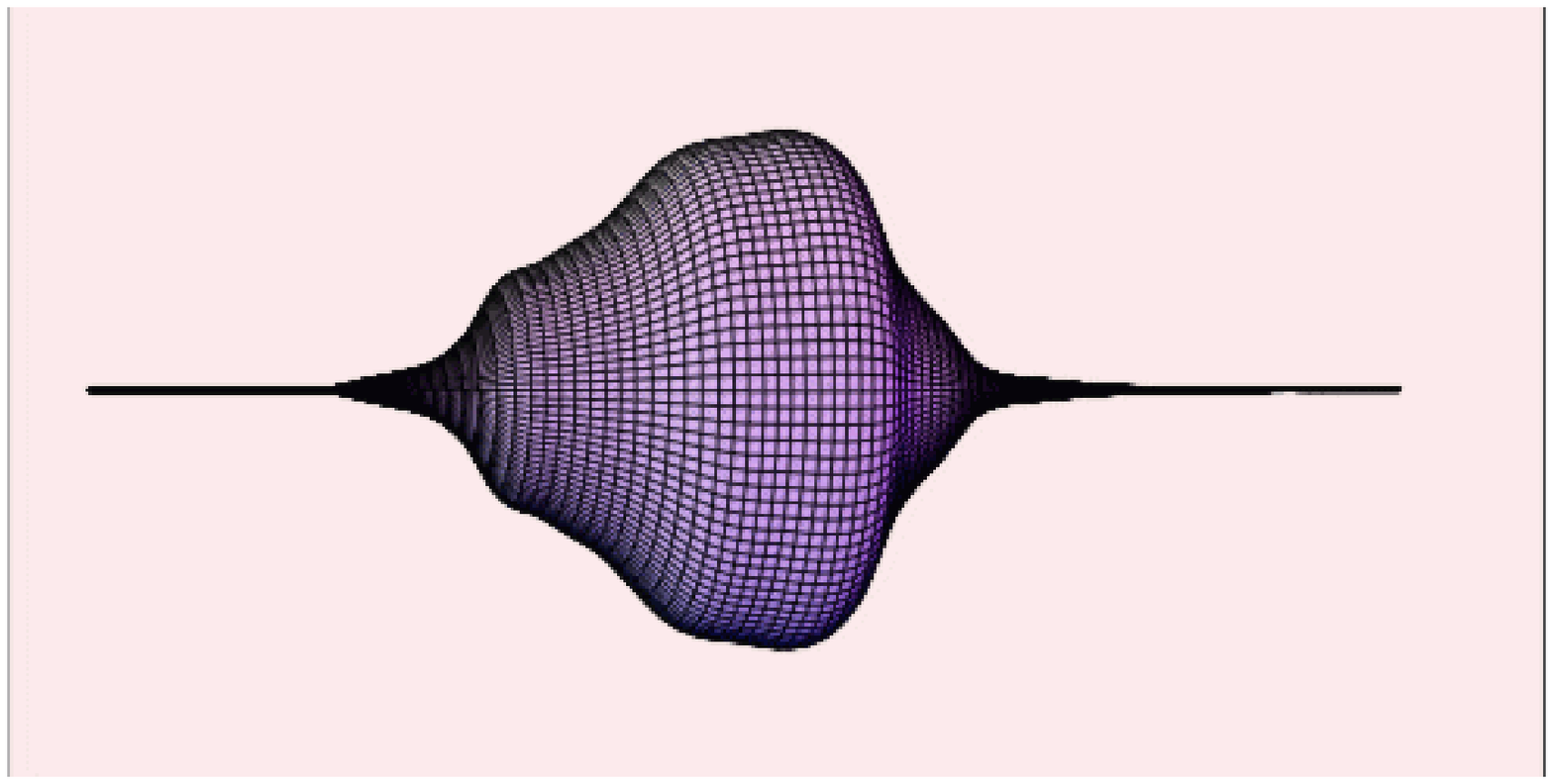}}}}}
%\end{flushleft}
\caption{The volume profiles of typical configurations in the phases 
A, B and C.
Phase C (bottom figure) is the one where extended four-dimensional 
geometries emerge.}
\label{blobs}
\end{figure}

We can now give the following qualitative characterization of the three
phases in terms of what we will provisionally call ``average geometry". 
The universe of phase C exhibits a classical four-dimensional background 
geometry on large scales, such that $\la {\it geometry}\ra \neq 0$. 
One may even argue that  $\la {\it geometry}\ra = const.$  
in view of the fact that according to the minisuperspace analysis of 
\cite{agjl,bigs4,semi} and allowing for a finite, global rescaling of the
renormalized proper time, the universe can be identified with the round 
$S^4$, a maximally symmetric de Sitter space of constant scalar curvature
(as we will describe in detail below). 
By contrast, in phase B the universe presumably has no extension or
trace of classicality, corresponding to $\la {\it geometry}\ra = 0$.
Lastly, in phase A, the geometry of the universe appears to be
``oscillating'' in the time direction. The different behaviour of
typical configurations is shown in Fig.\ \ref{blobs}. The ``time''
direction is horizontal and we plot the three-volume $N_3(t)$, i.e.\
the number of tetrahedra in a given time slice, as the
circumference of a circle. 
The three phases are separated by three phase transition lines which meet
in a triple point as illustrated in Fig.\ \ref{figphdiagram}.

\section{The macroscopic de Sitter universe (phase C)\label{S4}}  

\subsection{Identifying the infrared part of the universe}

Phase C in the above-mentioned phase diagram has our main interest,
because this is where we observe an extended four-dimensional universe. 
We will now discuss in more detail the geometric 
properties of this ``macroscopic'' 
universe. The measurements reported in this Section have been performed
at the values $(\k_0,\Del) = (2.2,0.6)$ of the bare coupling constants,
a point which lies well inside phase C.

The Monte Carlo simulations referred to above will generate a sequence of 
spacetime histories. An individual spacetime history is not an
observable, in the same way as a path $x(t)$ of a particle in the 
quantum-mechanical path integral is not. However, 
it is perfectly legitimate to talk about the 
{\it expectation value} $\la x(t) \ra$ as 
well as the {\it fluctuations around} $\la x(t) \ra$, both of which
are in principle calculable in quantum mechanics. 

Obviously, there are many more dynamical variables in quantum gravity than 
there are in the particle case. 
We can still imitate the quantum-mechanical situation 
by picking out a particular one, for example, 
the spatial three-volume $V_3(t)$ at proper time $t$. We 
can measure both its expectation value $\la V_3(t)\ra $ as well as fluctuations
around it. The former gives us information about the large-scale  
``shape'' of the universe we have created in the computer. 
A ``measurement'' of $V_3(t)$ consists of a table $N_3(i)$, where
$i=1,\ldots,N_t$ and $N_t$ denotes the total number of time slices. The 
time axis has a total length of $N_t$ time steps, where $N_t=80$ in the 
actual simulations, and we have cyclically identified time slice $N_t+1$ with
time slice 1. 

What we observe in the simulations is that for the range of discrete volumes
$N_4$ under study the universe does {\it not} extend
(i.e. has appreciable three-volume) over the entire time axis, but rather is
localized in a region much shorter than 80 time steps. 
Outside this region the spatial extension $N_3(i)$ will be minimal, 
consisting of the minimal number (five) of tetrahedra needed to 
form a three-sphere $S^3$, plus occasionally a few more 
tetrahedra.\footnote{This
kinematic constraint ensures that the triangulation remains a {\it simplicial
manifold} in which, for example, 
two $d$-simplices are not allowed to have more than 
one $(d-1)$-simplex in common.}
This thin ``stalk" therefore
carries little four-volume, which means that in a given 
simulation we can for most practical purposes
consider the total four-volume of the remainder, 
the extended universe, as fixed. 

\begin{figure}
\centerline{\scalebox{0.75}{\rotatebox{0}{\includegraphics{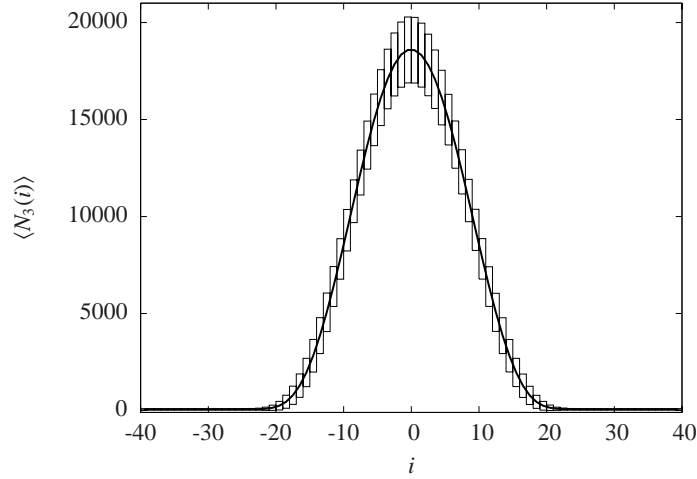}}}}
\caption{\label{fig1} Background geometry $\langle N_3(i)\rangle$: 
MC measurements for fixed $N_4=362.000$
and best fit \rf{n1} yield indistinguishable curves at given plot resolution. 
The bars indicate the average size of quantum fluctuations.}
\end{figure}

In order to perform a meaningful average over geometries 
which explicitly refers to the extended part of the universe, 
we have to remove the translational zero mode present, see 
\cite{bigs4} for a discussion of the procedure.
Having defined the ``centre of volume" along the 
time direction of our spacetime configurations, we can now 
perform superpositions of configurations and 
define the average $\la N_3(i)\ra $ as a function of the discrete time $i$.
The results of measuring this average discrete 
spatial size of the universe at various 
discrete times $i$ are illustrated 
in Fig.\ \ref{fig1} and can be succinctly summarized by the formula
\beq\label{n1}
N_3^{cl}(i):= \la N_3(i)\ra  = 
\frac{N_4}{2(1+\xi)}\;\frac{3}{4} \frac{1}{s_0 N_4^{1/4}}  
\cos^3 \left(\frac{i}{s_0 N_4^{1/4}}\right),~~~s_0\approx 0.59,
\eeq
where $N_3(i)$ denotes the number of three-simplices in the spatial slice 
at discretized time $i$ and $N_4$ the 
total number of four-simplices in the entire universe.
$\xi$ is a constant referring to the fact that we have a nonvanishing
asymmetry $\Del$, which implies
different lengths for time- and space-like links and consequently
different four-volumes for four-simplices of type (4,1) and (3,2). Likewise,
the ratio $N^{(4,1)}_4/N^{(3,2)}_4$ depends on the choice of 
bare coupling constants and has to be measured.   
Of course, formula \rf{n1} is only valid in the extended part of the universe 
where the spatial three-volumes are larger than the minimal cut-off size.

\begin{figure}[t]
\psfrag{t}{{\bf{\large $\sigma$}}}
\psfrag{v}{{\bf{\large $P(\sigma)$}}}
\centerline{\scalebox{1.1}{\rotatebox{0}{\includegraphics{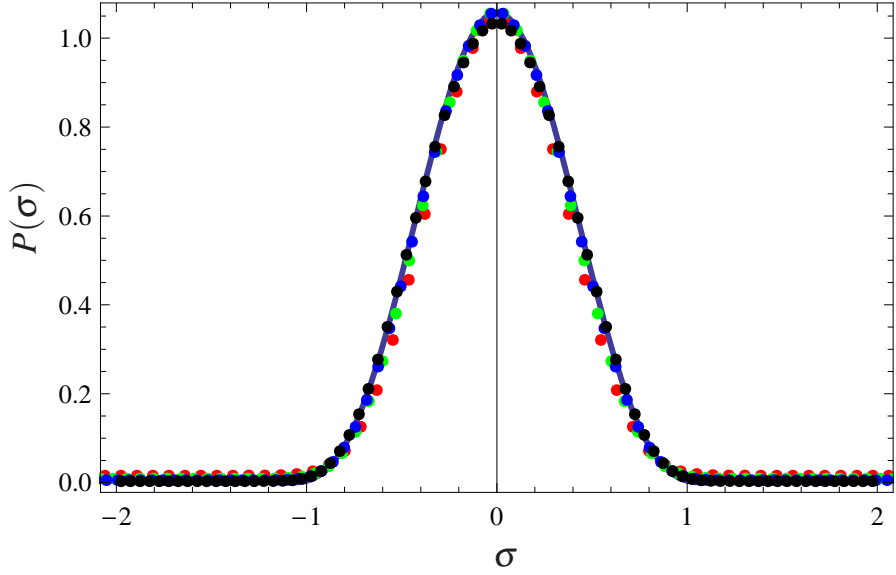}}}}
\caption{Rescaling of time and volume variables according to relation
\rf{n1} for $N_4 =$ 45.500, 91.000, 181.000 and 362.000. The plot 
also include the curve \rf{n1}. More precisely: $\sg \propto 
i/N_4^{1/4}$ and $P(\sg) \propto N_3(i)/N_4^{3/4}$.}
\label{fig1a}
\end{figure}

The data shown in Fig.\ \ref{fig1} have been 
collected at the couplings $(\k_0,\Del) = (2.2,0.6)$ and for 
$N_4= 362.000$. 
For these particular values of $(\k_0,\Del)$
we have verified relation \rf{n1} for $N_4$ ranging from 45.500 to 362.000
building blocks (45.500, 91.000, 181.000 and 362.000). 
After rescaling the time and volume variables by suitable powers of $N_4$ 
according to relation (\ref{n1}), and plotting them in the same way as 
in Fig.\ \ref{fig1}, one finds almost total
agreement between the curves for different spacetime volumes,
as illustrated in Fig.\ \ref{fig1a}. This constitutes
a beautiful example of finite-size scaling. At least with regard to 
measuring the average three-volume $V_3(t)$ all our discretized
volumes $N_4$ are sufficiently large to be treated as infinite, in the
sense that no further changes will occur for larger $N_4$.

By contrast, the quantum fluctuations indicated in 
Fig.\ \ref{fig1} as vertical bars for each discrete time $i$ {\it are} 
volume-dependent and will become (relatively) larger when the
the total four-volume is decreased.
Eq.\ \rf{n1} shows that
spatial volumes scale according to $N_4^{3/4}$ and time intervals
according to $N_4^{1/4}$, as one would expect for
a genuinely {\it four}-dimensional spacetime. This is exactly 
the scaling we have used in Fig.\ \ref{fig1a}. It strongly suggests
a translation of \rf{n1} to a continuum notation.
The most natural identification is given by  
\beq\label{n2}
\sqrt{g_{tt}}\; V_3^{cl}(t) = V_4 \;
\frac{3}{4 B} \cos^3 \left(\frac{t}{B} \right),
\eeq
where we have made the identifications
\beq\label{n3}
\frac{t_i}{B} = \frac{i}{s_0 N_4^{1/4}}, ~~~~
\Del t_i \sqrt{g_{tt}}\;V_3(t_i) = 2 \tC_4 N_3(i) a^4,
\eeq
such that
\beq\label{n3z}
\int dt \sqrt{g_{tt}} \; V_3(t) = V_4.
\eeq
In \rf{n3}, $\sqrt{g_{tt}}$ is the constant proportionality 
factor between the time
$t$ and genuine continuum proper time $\tau$, $\tau=\sqrt{g_{tt}}\; t$.
(The combination $\Del t_i\sqrt{g_{tt}}V_3$ contains $\tC_4$, related to the 
four-volume of a four-simplex rather than the three-volume corresponding
to a tetrahedron, because its time integral must equal $V_4$).
Writing $V_4 =8\pi^2 R^4/{3}$, and $\sqrt{g_{tt}}=R/B$,
eq.\ \rf{n2} is seen to describe 
a Euclidean {\it de Sitter universe} 
(a four-sphere, the maximally symmetric space for 
positive cosmological constant)
as our searched-for, dynamically generated background geometry!
In the parametrization of \rf{n2} 
this is the classical solution to the action
\beq\label{n5}
S= \frac{1}{24\pi G} \int d t \sqrt{g_{tt}}
\left( \frac{ g^{tt}\dot{V_3}^2(t)}{V_3(t)}+k_2 V_3^{1/3}(t)
-\lam V_3(t)\right),
\eeq
where $k_2= 9(2\pi^2)^{2/3}$ and  $\lam$ is a Lagrange multiplier,
fixed by requiring that the total four-volume 
be $V_4$, $\int d t \sqrt{g_{tt}} \;V_3(t) = V_4$. 
Up to an overall sign, this is precisely 
the Einstein-Hilbert action for the scale 
factor $a(t)$ of a homogeneous, isotropic universe
(rewritten in terms of the spatial three-volume $V_3(t) =2\pi^2 a(t)^3$), 
although we of course never put any such
simplifying symmetry assumptions into the CDT model.
The intriguing possibility of describing the data in terms of the minisuperspace 
model \rf{n5} was first reported in \cite{semi}. 

A discretized, dimensionless version of \rf{n5} is
\beq\label{n7b}
S_{discr} =
k_1 \sum_i \left(\frac{(N_3(i+1)-N_3(i))^2}{N_3(i)}+
\tilde{k}_2 N_3^{1/3}(i)\right),
\eeq
where $\tilde{k}_2\propto k_2$.
This can be seen by applying the scaling \rf{n1}, 
namely, $N_3(i) = N_4^{3/4} n_3(s_i)$
and $s_i = i/N_4^{1/4}$. 
This enables us to finally conclude that the identifications 
\rf{n3} when used in the action \rf{n7b} lead 
na\"ively to the continuum expression \rf{n5} under the 
identification\footnote{Due to the difference in four-volume 
between $N_4^{(3,2)}$ and $N_4^{(4,1)}$ for $\a \neq 1$ we have 
to introduce a compensating factor 
$\tilde{s}_0 \equiv s_0 \la N_4 \ra^{1/4}/\la N_4^{(4,1)}\ra^{1/4}$, see 
\cite{bigs4} for a detailed discussion.}
\beq\label{n7c}
G = \frac{a^2}{k_1} \frac{\sqrt{\tC_4}\; \ts_0^2}{3\sqrt{6} }.
\eeq
We also note that the reported scaling \rf{n1} implies that we can write
\beq\label{n7bb}
S_{discr} =
k_1 \sqrt{N_4} \;\sum_i \d s_i \;\left( 
\frac{1}{n_3(s_i)}\Big(\frac{n_3(s_{i+1})-n_3(s_i)}{\d s_i}\Big)^2+
\tilde{k}_2 n_3^{1/3}(s_i)\right),
\eeq
where $\d s_i= 1/N_4^{1/4}$. Thus, referring to \rf{ny17b}, 
we see that there are at least terms in $\log f(N_4,\k_0,\Del)$ which
scale like  $\sqrt{N_4}$. To obtain the form \rf{ny17b} we need 
additional terms of entropic nature since the coefficient 
of $\sqrt{N_4}$ in \rf{ny17b} is assumed positive.

\subsection{The size of the universe and the flow of $G$\label{size}}

It is natural to identify the coupling constant $G$ multiplying 
the effective action for the scale factor with the gravitational
coupling constant $G$. The effective action describing our 
computer-generated data is given by eq.\ \rf{n5}, and its
dimensionless lattice version by \rf{n7b}. The computer data 
allow us to extract $k_1 \propto a^2/G$, with $a$ the 
spatial lattice spacing, and the precise constant 
of proportionality given by eq.\ \rf{n7c}. 

For the bare coupling constants
$(\k_0,\Del)= (2.2,0.6)$ we have high-statistics measurements
for $N_4$ ranging between 45.500 and 362.000 four-simplices 
(equivalently, $N_4^{(4,1)}$
ranging between 20.000 and 160.000 four-simplices). The choice of 
$\Del$ determines the asymmetry parameter $\a$, and the 
choice of $(\k_0,\Del)$ determines the ratio $\xi$ 
between $N_4^{(3,2)}$ and $N_4^{(4,1)}$. This in turn determines 
the ``effective'' four-volume $\tC_4$ of an average four-simplex, which
also appears in \rf{n7c}. The number $\ts_0$ in \rf{n7c} 
is determined directly from the time 
extension $T_{\rm univ}$ of the extended universe according to
\beq\label{width}
T_{\rm univ}=\pi\; \ts_0 \Big(N_4^{(4,1)}\Big)^{1/4}.
\eeq
Finally, from our measurements we have determined $k_1= 0.038$. 
Taking everything together according to  \rf{n7c}, 
we obtain $G\approx 0.23 a^2$, or 
$\ell_{Pl}\approx 0.48 a$, where  $\ell_{Pl} = \sqrt{G}$ is the Planck length.

From the identification of the volume of the four-sphere, 
$V_4 =  8\pi^2 R^4/{3} = \tC_4 N_4^{(4,1)} a^4$,
we obtain that $R=3.1 a$. In other words, 
{\it the linear size $\pi R$ of the quantum de Sitter universes 
studied here lies in the range of 12-21 Planck lengths for $N_4$ in the 
range mentioned above and for the bare 
coupling constants chosen as $(\k_0,\Del)=(2.2,0.6)$}.

Our dynamically generated universes are therefore not very big, and the 
quantum fluctuations around their average shape are large, as is apparent from 
Fig.\ \ref{fig1}. The presence of such fluctuations is evident in the 
bottom snapshot picture of the extended universe shown in Fig.\ \ref{blobs}, 
whose volume profile deviates from that of a regular sphere. 
The point is of course that we have to perform an averaging process to obtain the
{\it expectation value} of the volume profile, and this is precisely
what we have been doing numerically. It is rather 
surprising that the semiclassical minisuperspace formulation gives an adequate
description -- at least for the volume profile -- for universes of such a small size, 
a fact that should be 
welcome news to anyone performing semiclassical calculations to describe the 
behaviour of the early universe.
However, when looking at more local geometric properties of the universe, 
our lattices are still coarse compared
to the Planck scale $\ell_{Pl}$ because the latter 
corresponds to roughly half a lattice
spacing. If we are after a theory of quantum gravity valid on all scales, 
we are specifically interested in uncovering phenomena 
associated with Planck-scale
physics. In order to collect data which are free from unphysical 
short-distance lattice artefacts at this
scale, we would ideally like to work with a
lattice spacing much smaller than the Planck length,
while still being able to set by hand the physical volume of the  
universe studied on the computer. The way to achieve this is 
by changing the bare coupling constants $\k_0,\Del$ such 
that the coefficient $K$ in \rf{n7c}, $G=: K a^2$, is changed to a 
larger value. However, $K$ is a combination of a number of factors and 
they might change differently when $\k_0,\Del$ are changed. It is 
thus a (computer-)experimental exercise to find a path in the $(\k_0,\Del)$ 
coupling-constant plane such that $K$ increases and we end up with 
$\ell_{Pl}= \sqrt{K} a \gg a$. We will discuss later whether such a 
path exists.

\section{Constructive evidence for the effective action\label{effective}}

We have found a perfect fit \rf{n1} to the emergent background 
geometry and the curve can be related to
the continuum effective action \rf{n5}. However, it is 
still of interest to investigate to what extent the action \rf{n7b} can be rederived
from the data. Interestingly, as we shall see below, this can largely be done.

The data at our disposal are: (i) the measurement of 
the three-volume $N_3(i)$
at the discrete time step $i$, and of 
the three-volume correlator $N_3(i) N_3(j)$. 
Having created $Q$ statistically independent 
configurations $N_3^{(q)} (i)$ by Monte Carlo 
simulation allows us to construct the average
\beq\label{5.1}
\bar{N}_3(i) := \la N_3(i) \ra \cong \frac{1}{Q} \sum_{q=1}^Q N_3^{(q)} (i),
\eeq
where the superscript in $(\cdot)^{(q)}$ denotes the result 
of the q'th configuration sampled; (ii) 
the covariance matrix
\beq\label{5.2}
C(i,j) \cong \frac{1}{Q} \sum_{q=1}^{Q} 
(N^{(q)}_3 (i) -\bar{N}_3(i))(N^{(q)}_3 (j) -\bar{N}_3(j)).
\eeq

We now assume we have a discretized action which can 
be expanded around the expectation value $\bar{N}_3(i)$
according to
\beq\label{5.6}
 S_{discr}[\bar{N}+n] = S_{discr}[\bar{N}] +
\oh \sum_{i,j} n_i \hat{P}_{ij} n_j  +O(n^3).
 \eeq
If the quadratic approximation describes 
the quantum fluctuations around the expectation value $\bar{N}$ well,
the inverse of the operator $\hat{P}$ will be a good 
approximation to the covariance matrix.
Conversely, still assuming the quadratic approximation gives 
a good description of the 
fluctuations, the $\hat{P}$ constructed from the covariance matrix will to a 
good approximation allow us to reconstruct the action via \rf{5.6}.

\begin{figure}[t]
\centerline{\scalebox{0.7}{\rotatebox{0}{\includegraphics{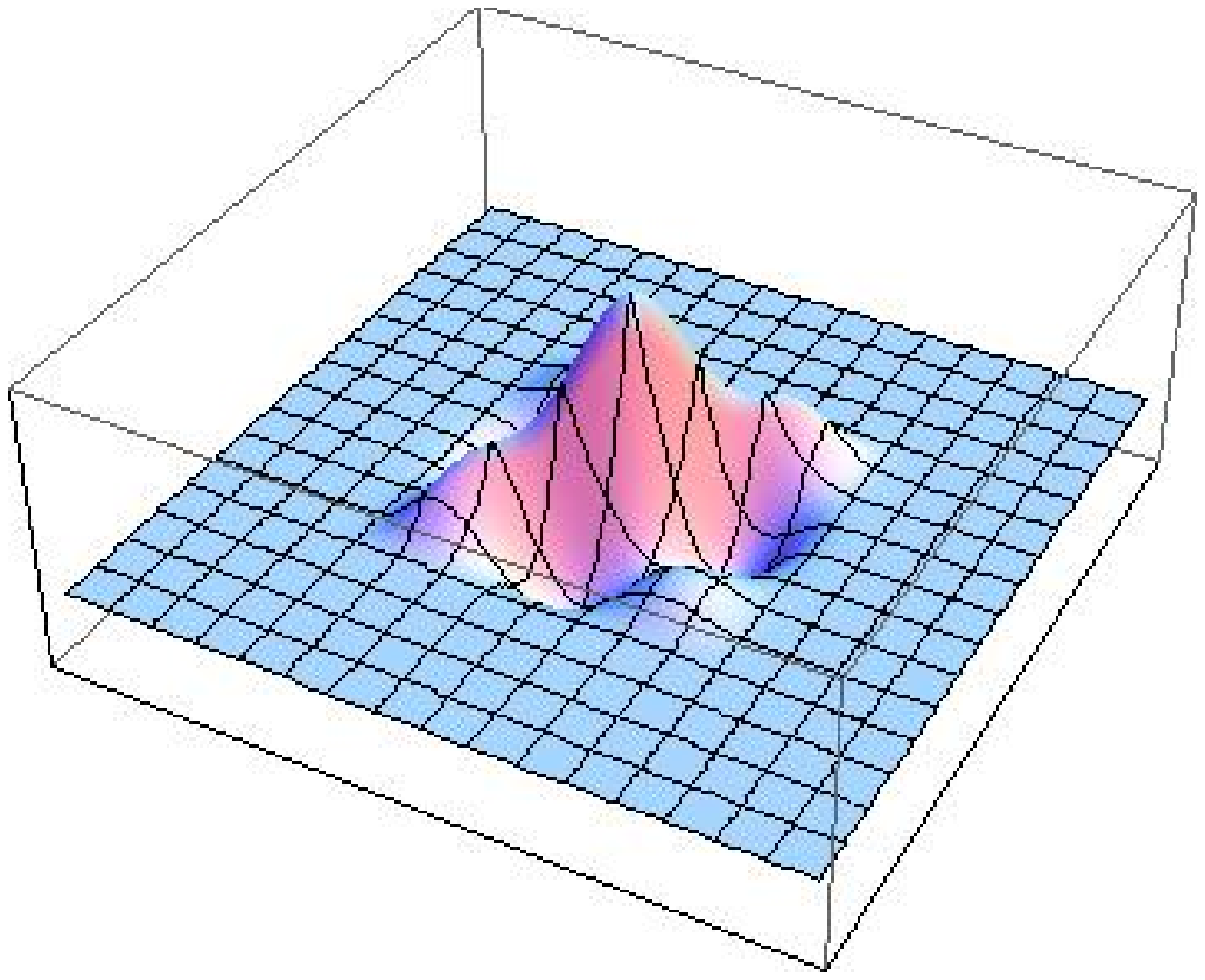}}}}
\centerline{\scalebox{0.7}{\rotatebox{0}{\includegraphics{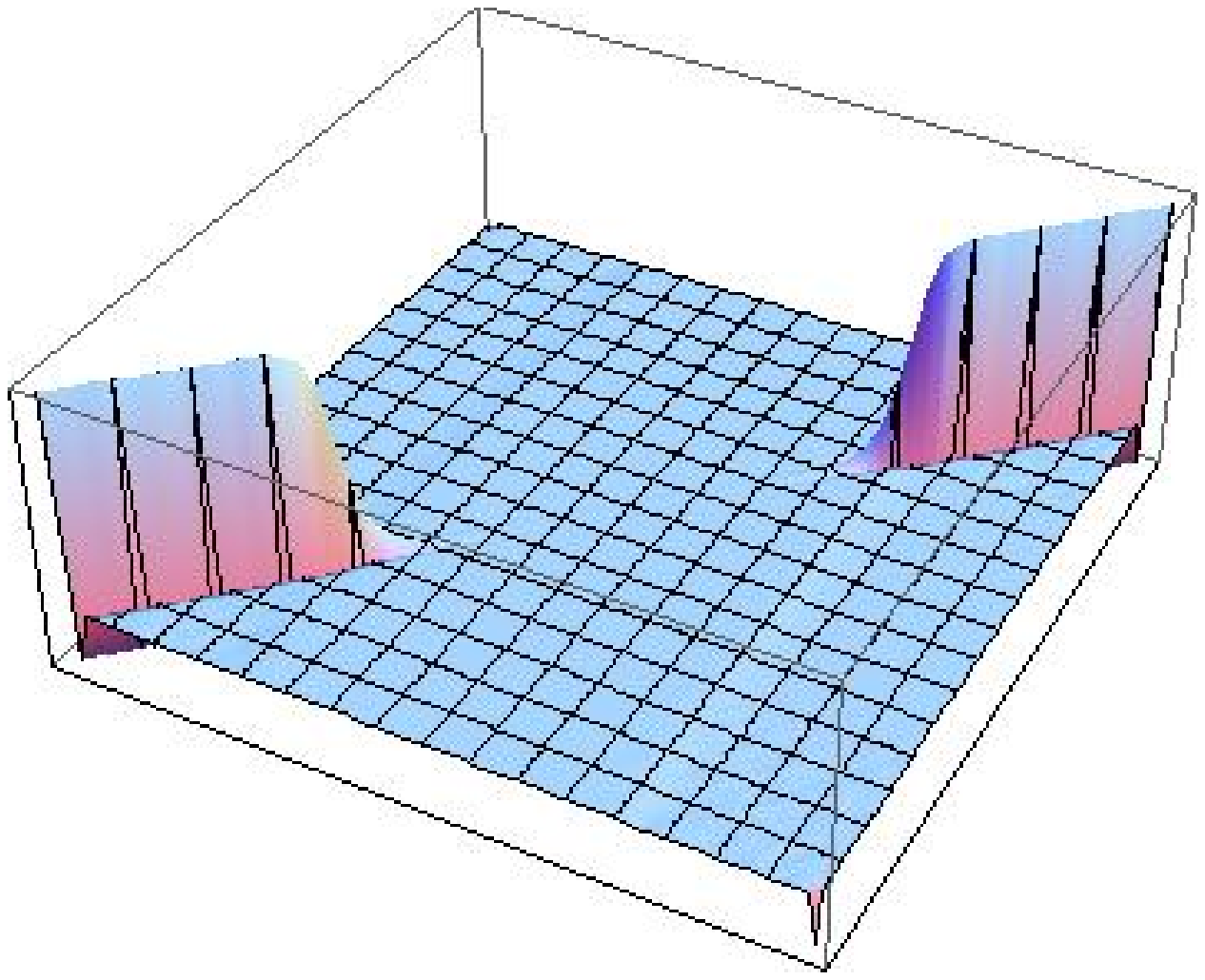}}}}
\caption{The covariance matrix $C$ (top) and its inverse (bottom).}
\label{propagator}
\end{figure}

Fig.\ \ref{propagator} shows the measured covariance matrix $C(i,j)$
and its inverse, the operator $\hat{P}$. Some care is needed in 
inverting $C(i,j)$ since it has two zero modes, one from the 
constraint that $N_4$ is kept fixed, and (an approximate) one 
from the fact that the translational mode of the 
`centre of volume' can only be fixed up to a lattice spacing,
see \cite{bigs4} for a detailed discussion.
As is clear from the figure, the inverse $\hat{P}$ is completely dominated
by the stalk data. This feature is unavoidable:
while the correlation matrix is dominated by long-range fluctuations,
the inverse matrix will be dominated by short-distance fluctuations,
i.e.\ the fluctuations in the stalk, which by definition
are associated with cut-off energies.

\begin{figure}[ht]
%\centerline{\scalebox{1.00}{\rotatebox{0}{\includegraphics{fig2.eps}}}}
\centerline{\scalebox{1.0}{\rotatebox{0}{\includegraphics{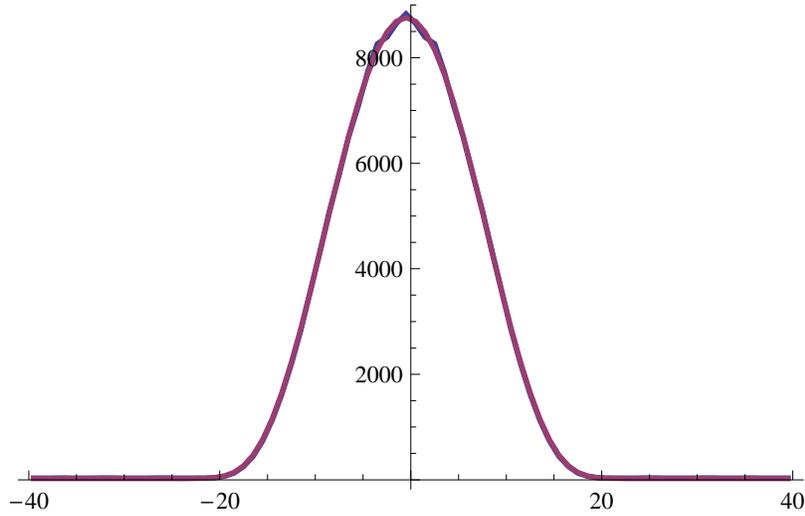}}}}
\caption{\label{fig4} The directly measured expectation values
$\bar{N}_3(i)$  compared to the averages $\bar{N}_3(i)$ 
reconstructed from \rf{5.11}, for $\k_0=2.2 $ and $\Del= 0.6$.}
\end{figure}

\begin{figure}[ht]
%\centerline{\scalebox{1.00}{\rotatebox{0}{\includegraphics{fig3.eps}}}}
\centerline{\scalebox{0.7}{\rotatebox{0}{\includegraphics{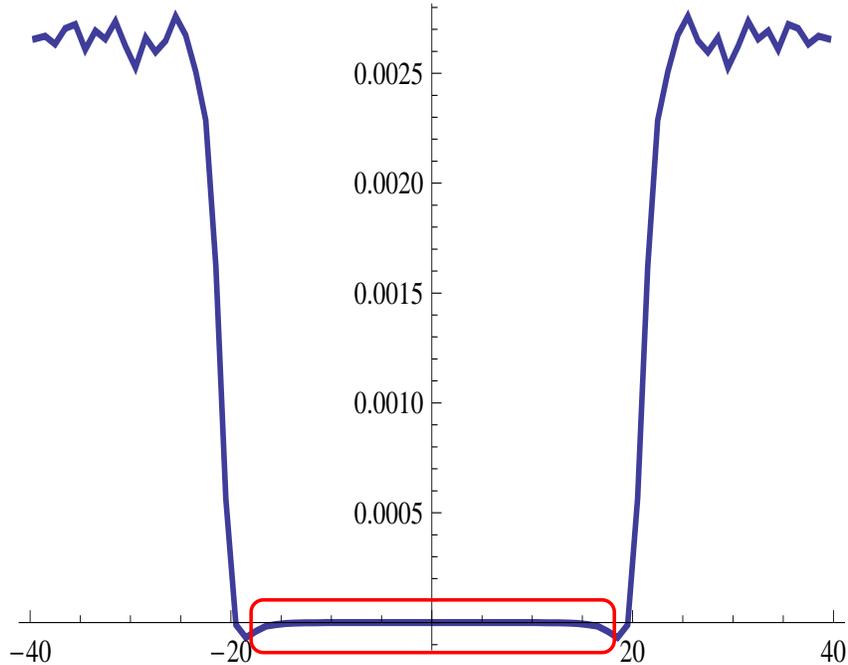}}}}
\caption{\label{fig4a} Reconstructing 
the second derivative $U''(\bar{N}_3(i))$ from
the coefficients $u_i$, for $\k_0=2.2$,
$\Del= 0.6$ and $N_4^{(4,1)}=160.000$. Data chosen from the
encircled region are independent of short-distance artefacts.}
\end{figure}

Looking at the inverse $\hat P$ of the measured covariance matrix, 
we observe that to very good approximation it is
small and constant, except on the 
diagonal and the entries neighbouring the diagonal. 
This means that we can decompose it into a ``kinetic'' and a ``potential'' term.
The kinetic part $\hPk$ is defined as the matrix with non-zero 
elements on the diagonal and in the 
neighbouring entries, such that the sum of the elements 
in a row or column is always zero. The potential part $\hPp$ is then 
given by whatever remains along the diagonal. 
We therefore arrive at a tentative representation
of $\hat{P}$ as
\beq\label{hatP}
\hat{P}_{ij} = \hPk_{ij} + \hPp_{ij},
\eeq
\beq\label{hatP1}
\hPk_{ij}=p_i \Del_{ij} ,~~~~~~\hPp_{ij}=u_i \del_{ij},
\eeq
where the matrices $\Del_{ij}$ and $\del_{ij}$ are essentially defined
through the construction just described\footnote{For
details of normalization and
subtleties in the definition of $\hPk$ and $\hPp$ related to the zero modes
we refer to \cite{bigs4}.}.
We know $\hat{P}$ from the data, and can make a
least-$\chi^2$ fit to determine the numbers $p_i$ and $u_i$.
For details we refer again to \cite{bigs4}. The results are
shown in Figs.\ \ref{fig4} and \ref{fig4a}. 

Let us look at the discretized minisuperspace action 
\rf{n7b} which has served as 
inspiration for the definition of $\hPk$ and $\hPp$. 
Expanding $N_3(i)$ to second order
around $\bN_3(i)$, one obtains the identifications
\beq\label{5.11}
\bN_3(i) = \frac{2 k_1}{p_i},~~~~~~U''(\bN_3(i)) = -u_i,
\eeq
where $U(N_3(i))=k_1\tilde k_2 N_3^{1/3}(i)$ 
denotes the potential term in \rf{n7b}. 
We use the fitted coefficients $p_i$
to reconstruct $\bN_3(i)$ and then compare these reconstructed 
values with the averages $\bN_3(i)$ measured directly. 
Similarly, we can use the measured $u_i$'s to 
reconstruct the second derivatives $U''(\bN_3(i))$ and 
compare them to the form $\bN^{-5/3}_3(i)$ coming from \rf{n7b}.

The reconstruction of $\bN_3(i)$ is illustrated in 
Fig.\ \ref{fig4} for a given 
four-volume $N_4$ and compared with the directly measured
expectation values $\bN_3(i)$. 
One observes that the reconstruction works very well and, 
most importantly, that the coupling constant $k_1$, which in this 
way can be determined independently for each 
four-volume $N_4$, really {\it is} independent of 
$N_4$ in the range of $N_4$'s we have considered, as it should be. 

We will now try to extract the potential $U''(\bN_3(i))$ from the 
information contained in the matrix $\hPp$. The determination of 
$U''(\bN_3(i))$ is not an
easy task as can be understood from Fig.\ \ref{fig4a}, 
which shows the measured coefficients $u_i$ extracted
from the matrix $\hPp$, and which we consider
rather remarkable. The interpolated curve makes an
abrupt jump by two orders of magnitude going from the 
extended part of the universe (stretching over roughly 40 time steps) 
to the stalk. The occurrence of 
this jump is entirely dynamical, since no distinction 
has ever been made by hand between stalk and bulk. 
In order to extract physical information related to 
a genuine potential like the one appearing in \rf{n7b},
we of course must restrict ourselves to the region inside the 
``blob'', corresponding to the data range 
encircled in Fig.\ \ref{fig4a}. From the figure it is also clear that
extracting $U''(\bN_3(i))$ from the 
data available is a nontrivial task.

\begin{figure}[ht]
\centerline{\scalebox{1.00}{\rotatebox{0}{\includegraphics{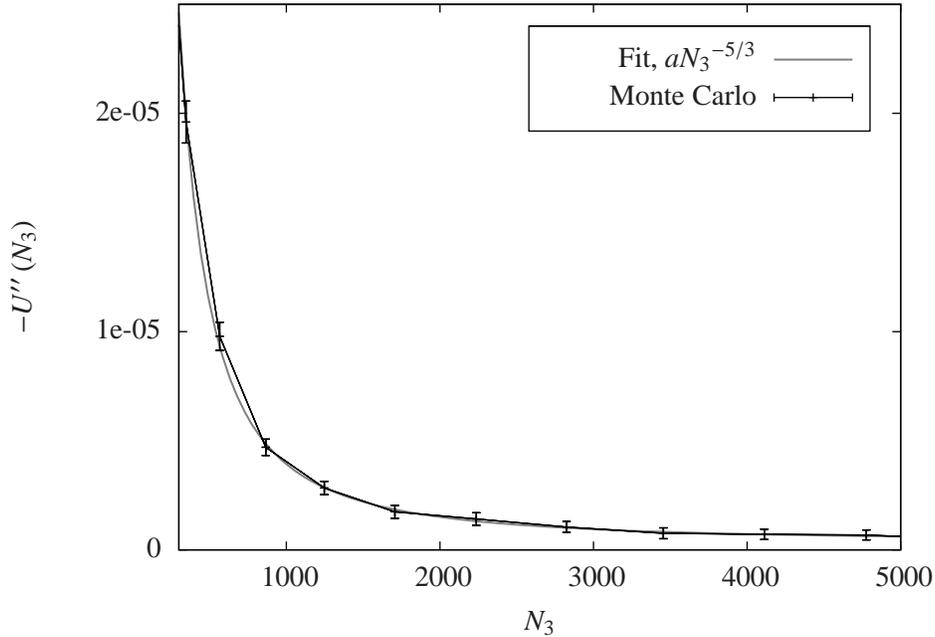}}}}
\caption{\label{fig5} The second derivative 
$-U''(N_3)$ as measured for $N_4^{(4,1)}= 160.000$,
$\k_0 = 2.2$ and $\Del = 0.6$.}
\end{figure}

The range of the discrete three-volumes $N_3(i)$ in the 
extended universe is from several thousand down to five, 
the kinematically allowed minimum. However, the behaviour 
for the very small values of $N_3(i)$ near the edge of the extended universe
is likely to be mixed in with 
discretization effects. 
In order to test whether one really has a $N_3^{1/3}(i)$-term
in the action one should therefore only use values of $N_3(i)$ somewhat larger
than five (shown as the encircled region in Fig.\ \ref{fig4a}). 
This has been done in Fig.\ \ref{fig5}, where we have 
converted the coefficients $u_i$ from functions of the discrete time steps $i$ 
into functions of the background spatial three-volume $\bN_3(i)$ 
using the identification in \rf{5.11} 
(the conversion factor can be read off the relevant curve
in Fig.\ \ref{fig4}). 
The data presented in Fig.\ \ref{fig5} were taken at a discrete volume
$N_4^{(4,1)}= 160.000$, and fit well the form $N_3^{-5/3}$, 
corresponding to a potential $\tk_2 N_3^{1/3}$. 

\begin{figure}[t]
\centerline{\scalebox{1.00}{\rotatebox{0}{\includegraphics{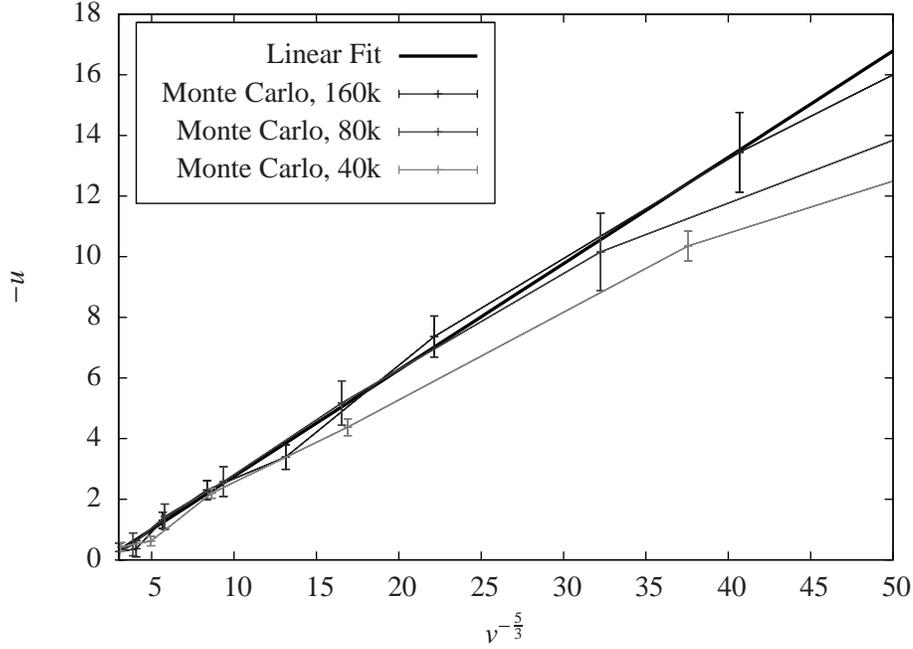}}}}
\caption{\label{fig6} The dimensionless 
second derivative $u= N_4^{5/4} U''(N_3)$ 
plotted against $\n^{-5/3}$, where $\n = N_3/N_4^{3/4}$ is the 
dimensionless spatial volume, for $N_4^{(4,1)}= 40.000$, 
80.000 and 160.000, $\k_0 = 2.2$ and $\Del = 0.6$. One expects a universal
straight line near the origin (i.e. for large volumes) if 
the power law $U(N_3) \propto N^{1/3}$ is correct.}
\end{figure}

Apart from obtaining the correct power $ N_3^{-5/3}$ for the potential
for a given spacetime volume $N_4$, 
it is equally important that the coefficient 
in front of this term be independent of $N_4$. 
This seems to be the case as is shown in Fig.\ \ref{fig6}, where
we have plotted the measured potentials in terms of reduced, dimensionless
variables which make the comparison between measurements for 
different $N_4$'s easier. --
In summary, we conclude that the data allow us to reconstruct the action
\rf{n7b} with good precision.

\section{Connection to Ho\v rava-Lifshitz gravity}

While we have verified that the action \rf{n7b} 
describes the data well, one aspect of the 
formula may lend itself to a more general interpretation. 
Until now we have 
chosen to view the ``experimental'' formula 
\rf{n1} as describing a round four-sphere 
via the identifications \rf{n2} and \rf{n3}.
However, the potential asymmetry between space and time introduced 
in our model by working with a time foliation allows for a different 
interpretation, namely, that space and time really behave differently,
as we will explain in what follows.

Although at the level of the 
piecewise linear structures we have a precise relation between 
the coupling constant $\Del$ and the asymmetry parameter $\a$,
this relation enters the construction only in a relatively weak way,
in that for a given $\Delta$ the bare action we use is
the Regge-Einstein-Hilbert action for a piecewise linear manifold 
with the given connectivity and the length assignment $a_t^2 = \a a_s^2$.
Nowhere else does this length assignment appear. Of course, 
{\it if} the model had a well-defined perturbative expansion,
one could have chosen the bare coupling 
constants such that the classical action (and, by implication, the
relation between $a_s$ and $a_t$) played an important role in the 
path integral. However, as already explained in the introduction, 
this is not the case. Rather,
our choice of bare coupling constants is dictated by the 
wish to find nontrivial 
physics, which restricts us to a region far away from where the 
action term dominates the entropy of configurations. 

In other words, the effective 
action we have obtained bears only indirect traces of the classical 
action put into the path integral. Similarly, the precise relation 
between space and time directions in the final continuum theory will
be determined by the statistical averages resulting from the full path integral, 
rather than the parameter $\a$ put into the discretized action. 

To illustrate the nonperturbative mechanism at work,  
let us consider the measurements in phase C, where we observe
a macroscopic universe whose extension in the time direction scales
like $N_4^{1/4}$ for different, fixed four-volumes $N_4$.
It should be emphasized that this scaling behaviour is by no means 
predetermined, for example, time in phase B scales completely differently
(in fact, the time extension vanishes there, which is of course an 
extreme situation).
One is therefore led to conclude that time and (the linear extension of) 
space scale identically
in phase C. This is corroborated by evidence that well inside phase C 
there seems to be a well-defined notion of a ``physical'' proper time extent
independent of $\a$, as one would have expected na\"ively \cite{rounds4}.

However, the fact that
the behaviour in phase B is very different leaves open the interesting
possibility that a nontrivial scaling relation between time and spatial
extent may ensue {\it when one approaches the B-C phase transition line}.
In that case one could try to describe the situation by a continuum effective 
action where time and space have different dimensionality. This is 
precisely what P.\ Ho\v rava attempted in his novel class of gravity
theories \cite{horava}, a topic we will return to shortly. 
But even in phase $C$, where time and 
space scale in the same way, one could explore the consequences of 
relaxing the notion that time and space should be related exactly as they are in 
general relativity. What we have done until now is to interpret our results in the 
continuum limit in terms of 
the classical theory. This can be achieved by making a global
rescaling of continuum proper time, which we have direct access to
through the preferred time foliation. More specifically,  
we have been checking our data against the cosmological minisuperspace 
model \rf{n5}, which can be derived from general relativity by assuming spatial 
homogeneity and isotropy. Eq.\ \rf{n5}, when written in terms of the 
scale factor\footnote{equivalently, one can also work with the associated 
three-volume $V(\tau) = 2\pi^2 a(\tau)^3$ as the basic configuration space
variable} $a(\tau)$, $\tau$ denoting 
proper time, reads
\beq\label{ny20}
S=\frac{3 \pi}{4} \int \d \tau \; \Big( a \dot{a}^2 +a -\lam a^3\Big).
\eeq
All our measurements are perfectly consistent with an effective
action of this form, but the ``cosmological'' observables we have 
been considering so far cannot discriminate between this and more
general cosmologies coming from a generalized ``gravity" theory with a built-in
anisotropy between space and time, like Ho\v rava's. In the latter, one can
again assume {\it spatial} homogeneity and
isotropy to obtain cosmological solutions,   
which in the Euclidean sector, {\it in the infrared limit}, arise 
from an action quite similar to \rf{ny20} \cite{elias,brandenberger,calcagni}, 
namely,
\beq\label{ny21} 
S = \frac{\pi}{8} \int \d \tau \; \Big(3(3\hat{\lam}-1) a \dot{a}^2+ 
6\g a - a^3(6 {\lam}+\tV(a)) \Big).
\eeq
The ``potential'' $\tV(a)$ in this expression has an expansion in inverse powers 
of $a$, coming from the higher-order spatial derivative terms
in the Ho\v rava-Lifshitz action. In the actual computer 
simulations, in both \rf{ny20} and \rf{ny21}
$\lam$ is a Lagrange multiplier rather than a 
cosmological constant, which  
ensures that the four-volume is kept fixed. As long as 
we can fix proper time only up to a constant and 
as long as we cannot measure reliably the correction term $\tV(a)$ containing inverse 
powers of $a$, it is not really possible 
to distinguish ``experimentally''  between  \rf{ny20} and \rf{ny21} in terms of a 
reconstruction of the action, as we did in the previous Section. 
In this situation, whenever
$\hat{\lam} > 1/3$ and $\g >0$, a rescaling of time in \rf{ny21} leads
to the same form as \rf{ny20} up to a constant of proportionality.
Taking into account the difficulties in verifying the mere existence
of the linear term in \rf{ny20} from the data (cf. Fig.\ \ref{fig4a}),
it is clear that we cannot presently extract from the data in a 
reliable way a potential $\tV(a)$ that depends on inverse powers of 
$a$, starting with $a^{-4}$. The only region in Fig.\ \ref{fig4a}
where such information could be extracted is for the smallest values of $a$.
Unfortunately, this is also where lattice artefacts will be important 
and artefacts from the ``stalk'' may get mixed in with genuine continuum
physics; a glance at Fig.\ \ref{fig4a} reveals how large the effects of the 
stalk are right next to the small-$a$ region.
  
This not withstanding, we can discuss the qualitative correspondence 
between the Ho\v rava scenario and our phase diagram.
In our earlier analysis of the different phases of CDT quantum gravity, we have 
chosen for a particular qualitative description to match precisely 
that of a Lifshitz phase diagram \cite{lifshitz,gold}.
The qualitative feature we want to emphasize in this context is that the role
played by ``average geometry'' in quantum gravity bears an intriguing resemblance to that
played by the Lifshitz field $\phi$.
In an effective Lifshitz theory, the Landau free energy density $F(x)$ 
as function of an order parameter $\phi(x)$
takes the form\footnote{see, for example, \cite{gold} for an introduction to
the content and scope of ``Landau theory"} 
\beq\label{2.2}
F(x) = a_2 \phi(x)^2 + a_4 \phi(x)^4 +a_6\phi(x)^6 + \ldots 
+c_2(\prt_\a \phi)^2 +d_2 (\prt_\b \phi)^2
+ e_2 (\prt_\b^2 \phi)^2 +\ldots ,
\eeq
where for a $d$-dimensional system $\a =m+1,\ldots,d$, $\b=1,\ldots,m$.
Distinguishing between ``$\alpha$"- and ``$\beta$"-directions allows
one to take anisotropic behaviour into account.
For a usual system, $m=0$ and a phase transition can occur when 
$a_2$ passes through zero (say, as a function of temperature).
For $a_2> 0$ we have $\phi=0$, while for $a_2 <0$ we have $|\phi| >0$ 
(always assuming $a_4 >0$).  However, one also has a transition
when anisotropy is present ($m>0$) and 
$d_2$ passes through zero. For negative $d_2$
one can then have an oscillating behaviour of $\phi$ in the $m$ 
``$\beta$"-directions.
Depending on the sign of $a_2$, the transition to this 
so-called modulated or helical phase can occur either from the 
phase where $\phi=0$, or from the phase where $|\phi| >0$. 
We conclude that the phases C, B, and A 
of CDT quantum gravity depicted in
Fig.\ \ref{fig1} can be put into one-to-one correspondence 
with the ferromagnetic, paramagnetic and helical phases
of the Lifshitz phase diagram\footnote{For 
definiteness, we are using here a ``magnetic'' language for the Lifshitz 
diagram. However, the Lifshitz diagram can 
also describe a variety of other systems, for instance, liquid crystals.}
if we in place of $\phi$ we use ``average geometry''. 
The triple point
where the three phases meet is the so-called Lifshitz point, where
in the Lifshitz model one can have a nontrivial scaling.

The critical dimension beyond which the mean-field 
Lifshitz theory alluded to above is believed to be valid is 
$d_c = 4+m/2$. In lower dimensions, the fluctuations play an
important role and so does the number of components of the field $\phi$.
This does not necessarily affect the general structure of the phase
diagram, but can alter the order of the transitions.
Without entering into the details of the rather complex general situation,
let us just mention that for $m=1$ fluctuations will often turn the
transition along the A-C phase boundary into a first-order transition.
Likewise, most often the transition between phases B and C is 
of second order. 
 
We conclude that the structure of the Lifshitz phase diagram 
is presently compatible with our CDT observations. Based on the order
parameter investigated in \cite{horava-simu}, 
the A-C transition of CDT quantum gravity looks like a clear-cut 
first-order transition.
On the other hand, the verdict is still out for 
the B-C line. The signals from the Monte Carlo simulations 
are ambiguous, and it appears that close to the transition line 
our algorithms
for updating the geometries need to be improved to produce 
fully reliable results. 

Phase transitions of higher than first order are of intrinsic interest
since they may serve as points where one can define the continuum 
theory. Since it appears that we have a well-defined 
infrared limit in phase C, such points would naturally 
be UV fixed points, and moving away from them
should bring us to the IR limit. Given the overall structure of the phase
diagram, the possible scenarios are as follows:
if either the A-C or the B-C line is second-order, we can 
potentially use any point on it to attempt
to define a UV limit. By contrast, if they are first-order lines, we are 
left with two interesting points which may be associated with a 
higher-order transition: the endpoint $P_0$ of the open B-C line
(cf. Fig.\ \ref{figphdiagram}), 
where the phase transition may be of higher order than along the line itself,
and the Lifshitz triple point $P_t$, where the transition may also be 
of higher order. 

Assuming there were such higher-order phase transition points,
how would we determine whether the UV limit of the theory is 
isotropic in space and time or, more generally, of Ho\v rava-type?
One defining aspect of Ho\v rava-Lifshitz gravity is the assumption that 
the scaling dimensions of space and time differ in the 
ultraviolet regime. This difference is used to construct a theory 
containing higher-order spatial derivatives in such a way that it
is renormalizable. How would one observe such a difference 
in the present lattice approach? -- Consider a universe of time 
extent $T$, spatial extension $L$ and total four-volume $V_4(T,L)$.
By measuring $T$ and $L$ we can establish the mutual relations
\beq\label{5.1a}
T \propto V_4^{1/d_t},~~~~L\propto \Big(V_4^{1-1/d_t}\Big)^{1/d_s}\propto 
T^{(d_t-1)/d_s}.
\eeq
Well inside phase $C$ we have measured $d_t=4$ and $d_s=3$, in agreement
with what is expected for an ordinary four-dimensional spacetime.
If the dimension [T] of time was $z$ times the dimension [L] of length,
we would have
\beq\label{5.2a}
z= \frac{d_s}{d_t-1}.
\eeq
We observed earlier that well inside phase B both $d_s$ and $d_t$ must
be large, if not infinite. In case the B-C phase transition is second-order, it  
may happen that $z$ goes to a value different from 1 when we approach
the transition line. To investigate this possibility, we have tried to 
determine $z$ as a function 
of the parameter $\Del$ as $\Delta\rightarrow 0$.
For $\Del > 0.3$ one obtains convincingly $d_t \approx4$ and 
$d_s \approx 3$ and thus $z\approx 1$. We can make an even stronger
statement, namely, that the data does not contradict the interpretation of 
$\Del$ as an (unphysical) asymmetry parameter: when $\Del$ is increased the 
corresponding $\a$ is decreasing (see Fig.\  \ref{alfa}), while the 
number of lattice spacings in the time direction is increasing, which 
at least qualitatively allows for the interpretation 
that the physical ``time'' is independent of $\a$.
By contrast, for $\Del \lesssim 0.3$ 
the quality of our results does not allow for any definite statements. 
Autocorrelation times become very long and there may be large 
finite-volume effects, which obscure the measurements and which are 
precisely based on finite-size scaling. 

To summarize, there is still a distinct
possibility that a nontrivial scaling with $z \neq 1$ can occur when 
one approaches the transition line between phases B and C. 
Should the B-C transition turn out to
be a first-order transition, the interesting points
would be $P_0$ and $P_t$. It would then be natural to conjecture 
that the point $P_t$ is a Ho\v rava-Lifshitz point with a nontrivial
scaling relation between space and time. The point $P_0$, which does
not appear in a standard Lifshitz diagram, would perhaps be a natural
candidate for an isotropic scaling point. On the other hand, if the 
B-C transition line is second-order,  
it leaves open the interesting possibility that the critical exponent 
changes continuously from $z=3$ at the Lifshitz point $P_t$ to $z=1$ at the 
(hypothetically) isotropic point $P_0$.

\section{Making contact with asymptotic safety}

As we discussed earlier, it is presently difficult to get close to the 
B-C phase transition line, which is needed if we want to achieve a
higher resolution in the UV regime, such that the lattice spacing 
is much smaller than the Planck 
length. Also, we do not know yet whether in such a limit
we have isotropy in space and time, like in the asymptotic safety approach,
or need to invoke an anisotropic scenario as outlined above. 
For the time being, let us assume that the endpoint 
$P_0$ of the B-C transition line in the phase diagram of Fig.\ \ref{figphdiagram} 
corresponds to an isotropic phase transition point.
How can one make contact with the gravitational
renormalization group treatment?
The standard way would be to ``measure'' observables (by lattice Monte 
Carlo simulations), like a mass 
in QCD or the string tension in Yang-Mills theory. 
For definiteness, consider the string tension, which
has mass dimension two. The measurements, for some choice $g_0$ of 
the bare coupling constant, will give us a number $\sigma(g_0)$.
We now write 
\beq\label{ny40}
\sigma(g_0) = \sigma_R a^2(g_0),
\eeq
where $\sigma_R$ is the physical string tension and $a(g_0)$ describes
the dependence of the lattice spacing $a$ on the bare coupling constant.
Being able to write down a relation like this for 
all observables, where $a(g_0)$ is determined by the renormalization
group equation
\beq\label{ny41}
a \frac{\d g_0}{\d a} = -\b (g_0),
\eeq
allows us to define a continuum theory at a fixed point $g_0^*$ where 
$\b(g_0)=0$, since there we can take $a(g_0) \to 0$ when $g_0 \to g_0^*$.
In the case of QCD or Yang-Mills theory the fixed point is the Gaussian
fixed point $g_0^*=0$, but in the more general setting of asymptotic safety
it will be non-Gaussian.

Assume now that we have a fixed point for gravity. The gravitational coupling 
constant is dimensionful, and we can write for the bare coupling
constant
\beq\label{ny42}
 G(a) = a^2 \hG(a),~~~~ a \frac{\d \hG}{\d a} = -\b(\hG),~~~~
\b(\hG) = 2\hG -c \hG^3 +\cdots\ .
\eeq   
The IR fixed point $\hG =0$ corresponds to $G$ constant while the 
putative non-Gaussian fixed point corresponds to $\hG\to \hG^*$, 
i.e.\ $G(a) \to 
\hG^* a^2$. In our case it is tempting to identify our dimensionless 
constant $k_1$ with $1/\hG$, up to the constant of proportionality 
given in \rf{n7c}. Close to the UV fixed point we have 
\beq\label{ny43}
\hG(a) = \hG^* - K a^\tc,~~~k_1 = k_1^* + K a^\tc,~~~~~~\tc=-\b'(\hG^*).
\eeq
Usually one relates the lattice spacing near the fixed point 
to the bare coupling constants with the help of some correlation
length $\xi$. Assume that $\xi$ diverges according to
\beq\label{ny44}
\xi(g_0) = \frac{c}{|g_0 -g_0^*|^\n}
\eeq
in the limit as we tune the bare coupling constant $g_0 \to g_0^*$.
This correlation length is associated with a field correlator and 
usually some physical mass $m_{ph}$ by means of
\beq\label{ny45}
\frac{|n_1-n_2|}{\xi(g_0)} = m_{ph} (a |n_1-n_2|) = m_{ph}|x_1-x_2|,
\eeq
where $|n_1-n_2|$ is a discrete lattice distance and 
$|x_1-x_2|$ a physical distance. Requiring the physical
quantities $|x_1-x_2|$ and 
$m_{ph}$ to remain constant as $a \to 0$ then fixes $a$ as a function
of the bare coupling constant,
\beq\label{ny46}
a = \frac{1}{c \, m_{ph}} \; |g_0-g_0^*|^\n. 
\eeq
Eq.\ \rf{ny46} is only valid close to the fixed point and 
should be compared to the renormalization group equation \rf{ny41},
from which we deduce that $\n = -1/|\b'(g_0^*)|$. 

In the gravitational case at hand we do not (yet) have  
observables which would allow
us to define meaningful correlation lengths. 
At any rate, it is by no means a settled issue 
how to {\it define} such a concept in a theory where one integrates 
over all geometries, and where the length is itself a function
of geometry (see \cite{correlationlength} for related discussions). 
Instead, we construct from our computer-generated ``data'' 
an effective action, where all degrees of freedom, 
apart from the scale factor, have been integrated out.  
We impose the constraint that the data are fitted 
to a universe of total lattice four-volume $N_4$. 
Measurements are performed at
different, fixed values of $N_4$, all the while maintaining the 
relation\footnote{in principle, we should be taking
into account the different volumes of the two types of
four-simplices, which depend on $\Del$, 
but we will ignore these details to streamline the presentation} 
\beq\label{ny47} 
 V_4 = N_4 a^4.
\eeq
We then ``remove the regulator", by investigating the limit 
$N_4\rightarrow\infty$.
In ordinary lattice field theory, we have two options for changing 
$N_4$; either we keep $a$ fixed, and therefore change $V_4$, or 
we keep $V_4$ fixed and change $a$. Let us illustrate the
difference in terms of a scalar field on a lattice.
Its dimensionless action can be written as 
\beq\label{ny48}
S= \sum_{i} \Big( \sum_\m (\phi(i+\m)-\phi(i))^2 + m_0^2  \phi^2(i)\Big),
\eeq
where $i$ labels discrete lattice points and $\m$ unit
vectors in the different lattice directions. The correlation length is 
approximately $1/m_0$ lattice spacings. Holding $a$ fixed and
increasing $N_4$ is not going to 
change the correlation length in any significant way if $N_4$ is sufficiently
large. Thus the interpretation for fixed $a$ is straightforward:
the physical volume $V_4$ is simply increased and finite-size 
effects will become smaller and smaller. 
However, we can also insist on an interpretation where $V_4$ is kept fixed,
$N_4$ is increased
and $a$ decreased accordingly. In this case, the lattice becomes finer
and finer with increasing $N_4$. But now the physical interpretation of 
\rf{ny48} will change with increasing $N_4$, even if no bare coupling 
constant is changed, and the correlation length is still approximately 
$1/m_0$ lattice spacings. Since the {\it physical} lattice length 
$a$ decreases proportional to $1/N_4^{1/4}$ the {\it physical} correlation
length is going to zero, and the physical mass to infinity. This can 
be made explicit in \rf{ny48} by introducing the lattice
spacing $a$,
\beq\label{ny49}
S= \frac{1}{a^2} \sum_i a^4 \left( \sum_\m 
\Big( \frac{(\phi(i+\m)-\phi(i))^2}{a^2}\Big)
+ \frac{m_0^2}{a^2}  \phi^2(i)\right).
\eeq
The physical mass is $m_{ph}=m_0/a$ and goes to infinity 
unless we adjust $m_0$. The factor $1/a^2$ in front of 
the sum can be reabsorbed in a redefinition of $\phi$ if desired.

In our case it is natural to consider $V_4$ as fixed if we want
to make contact with the continuum framework of asymptotic safety, 
since this will allow
us to vary $a$. Suppose we have identified a fixed point which 
we consider interesting, e.g.\ the point $P_0$
in our phase diagram. We now approach this point and measure 
$k_1$. If it is a UV fixed point, eq.\ \rf{ny43} tells us what to 
expect. Using \rf{ny47}, we can now convert this into an equation 
involving $N_4$, and suitable for application in CDT simulations,
\beq\label{ny50}
k_1(N_4)=k_1^c -\tilde{K} N_4^{-\tc/4}.
\eeq
When we measured $k_1(N_4)$ deep inside phase $C$ (at the
point ($\k_2,\Del)=(2.2,0.6)$),
we did not find any $N_4$-dependence of $k_1$. However, 
according to the insights just presented, 
we should observe such a dependence
at or close to a UV fixed point. As already noted earlier, an explicit
verification of such a relation will have to await more reliable 
computer simulations close to the phase transition lines. 

In fact, we have already 
seen indications in CDT quantum gravity of a short-distance behaviour
like that occurring in the asymptotic safety scenario. Recall the ``na\"ive''
renormalization conditions \rf{ny17a} and \rf{ny17d}. They were
introduced mainly to illustrate how a renormalization procedure could
lead to finite renormalized cosmological and gravitational constants, 
both with a semiclassical 
interpretation. If we are close to the UV fixed point, we know that 
$G$ will not be constant when we change scale, but $\hat G$ will.
Writing $G(a)= a^2 \hat G^*$, eqs.\ \rf{ny17a} and \rf{ny17d} are changed to
\beq\label{ny51}
\k_4-\k_4^c = \frac{\La}{\hat G^*} \,a^2,~~~~k_1(\k_0^c) = \frac{1}{\hat G^*}.
\eeq
The first of these relations now looks two-dimensional (cf. eq.\ \rf{ny4})!
Nevertheless, the expectation value of the four-volume still satisfies the 
correct relation
\beq\label{ny52}
\la V_4\ra = \la N_4\ra \;a^4 \propto \frac{1}{\La^2},
\eeq
as follows from \rf{ny17e}.

\begin{figure}[t]
%\vspace{-3cm}
\psfrag{X}{{$\;\;\;\sg$}}
\psfrag{Y}{{ $\!\!\! D_S$}}
\centerline{\scalebox{1.2}{\rotatebox{0}{\includegraphics{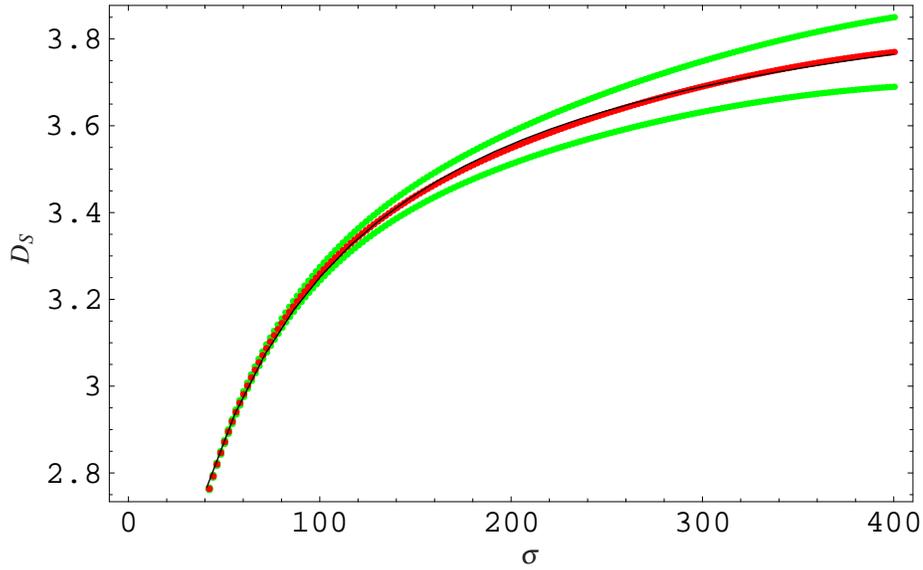}}}~~~~
~~~}
%\vspace{-1cm}
\caption[phased]{{\small The data points along the central curve show the
spectral dimension $D_S(\sg)$ of the universe as function
of the diffusion time $\sg$. Superimposed is a best fit, the continuous curve
$D_S(\sg) = 4.02\mi 119/(54\pl\sg)$. 
The two outer curves quantify the error bars, which
increase linearly with $\sg$.
(Measurements taken for a quantum universe with 181.000
four-simplices.) }}
\label{d4s2.2b4}
\end{figure}

Further hints of a two-dimensional signature at short distances have
come from measuring the so-called spectral dimension.
Essentially, this is the dimension a diffusing ``liquid" would 
experience in a spacetime with a typical quantum geometry of the kind 
appearing in the gravitational path integral, see 
the original article \cite{spectral}
for details. Fig.\ \ref{d4s2.2b4} recalls the result of measuring
the spectral dimension, which appears to change nontrivially as a 
function of diffusion time.    
Since short diffusion times probe 
short distances, we can read off from the fit indicated that the 
short-distance
spectral dimension is close to two. What is most intriguing is that the
short-distance result 
$D_S=2$ has since been found in two other quantum field-theoretic
approaches, namely, the renormalization group treatment \`a la Reuter
\cite{laureu} and Ho\v rava-Lifshitz gravity \cite{horspec}. 
At the same time, it means that this particular observable 
cannot distinguish between isotropic and anisotropic quantum gravity theories.

\section{Discussion\label{discussion}}

The CDT model of quantum gravity, which we have described in these lectures
is extraordinarily simple. It implements the path integral 
over causal geometries with a global time foliation. In order
to perform the integration explicitly, we introduced a grid of piecewise linear 
geometries, similar to how one proceeds when defining the path integral in 
ordinary quantum mechanics. Next, we rotated each of these geometries to Euclidean 
signature and used as our bare action the 
Einstein-Hilbert action in Regge form. In terms of ingredients, that's all.

The resulting superposition exhibits a 
nontrivial scaling behaviour as function of the 
four-volume, and we observe the appearance of a well-defined average 
geometry, that of de Sitter space, the maximally 
symmetric solution to the classical
Einstein equations in the presence of a positive cosmological constant.  
The measurements performed so far definitely probe the quantum regime,
since the fluctuations of the three-volume around de Sitter 
space are sizable, as can be seen in 
Fig.\ \ref{fig1}. Both the average volume profile and the quantum fluctuations 
around it are well described in terms of the mini\-superspace action \rf{n5}. 
A key feature to appreciate here is that, unlike in 
standard (quantum-)cosmological treatments, 
this description is the {\it outcome} of a nonperturbative evaluation
of the {\it full} path integral, 
with everything but the scale factor (equivalently, $V_3(t)$)
summed over. Measuring the correlations of the quantum fluctuations 
in the computer simulations for a particular choice of bare coupling constants
enabled us to determine the continuum gravitational coupling
constant $G$ as $G\approx 0.42 a^2$, thereby introducing an absolute physical 
length scale into the dimensionless lattice setting. 
Within measuring accuracy, our de Sitter universes (with volumes lying
in the range of 6.000-47.000 $\ell_{Pl}^4$) are seen to behave perfectly 
semiclassically with regard to their large-scale properties. 

These semiclassical results ``emerge'' even if, as emphasized above, we
are far from a region in coupling-constant space where the classical action
can be considered as dominant in the path integral. The resulting quantum geometry
comes about through the interplay of the weight provided by the 
exponential of the bare action and the weight provided by the entropy of a 
particular kind of configuration. From this point of view the results are
truly nonperturbative, even if they bear some similarity to the semiclassical
minisuperspace results. It should also be emphasized that we 
have only derived an effective action for the scale factor, not for 
the ``real'', transverse degrees of freedom. The issue of how to 
do this for the latter remains to be addressed.

The results we have reported are mostly ``infrared'' in nature. Our dynamically
generated universes
are macroscopic (although small) and -- with the exception of 
the spectral dimension measurements -- we are not yet probing Planckian 
(and possibly sub-Planckian) scales. 
It is a major issue whether such a short-distance 
completion of gravity exists in a conventional, field-theoretical
sense. ``Asymptotic safety" is an attempt to formulate the 
general conditions for this to be the case. In its simplest realization on the lattice
it requires a UV fixed point. This is precisely the kind of situation the CDT framework 
allows us to address; we have a phase diagram and 
potential fixed points, and as outlined above it is in principle
possible to check whether the UV behaviour is in accordance with the 
predictions from asymptotic safety. Critical slowing 
down close to the potential fixed points, i.e.\ a vast increase in the computer time 
needed to generate statistically independent geometries, 
has so far made it impossible to obtain reliable results, but work is now 
in progress to improve the updating algorithms.

Once we succeed in approaching the phase transition
line B-C in a reliable fashion, we may also be able to check 
whether the Ho\v rava-Lifshitz scenario is realized in our model. 
In addition to the standard asymptotic safety picture, the CDT construction
has the potential to accomodate also this scenario.
Its obvious similarities with the defining properties of 
Ho\v rava's ``anisotropic gravity" approach are the
distinguished role of a time foliation, and the presence of a unitary time
evolution (in CDT linked to the reflection positivity of the transfer matrix).

In addition to this, we have already pointed to the striking similarity between the
CDT and Lifshitz phase diagrams upon (somewhat loosely) 
identifying the Lifshitz mean-field
order parameter $\phi$ with ``average geometry".
If more specifically we want to relate the Lifshitz field to a mode
of the gravitational field, the conformal factor appears as a natural candidate.
The conformal mode is already known to play a decisive role in a variety of
geometro-dynamical contexts.
In the bare Euclidean Einstein-Hilbert action the kinetic term associated with the
conformal factor appears with the wrong, negative sign, leading to ill-defined
expressions for na\"ive cosmological, Euclidean path integrals.
In Euclidean noncritical
string theory the dynamics of the conformal factor is believed to cause a transition from
a ``healthy phase" (where $c\leq 1$ for the central charge of matter) to a
degenerate phase of so-called branched polymers (for $c>1$)\footnote{In 
noncritical string theory there exists an 
analytic proof that for the dimension of (Euclidean) 
target spacetime $d \geq 2$ the string surfaces degenerate into
branched polymers \cite{ad}, see also \cite{dfj} for similar results on a
hypercubic lattice.}. A similar
phenomenon was observed in the old Euclidean dynamical triangulations
approach to quantum gravity, where the bare (inverse) gravitational coupling
$\kappa_0$ plays the role of the central charge $c$, in the sense that for
large values of $\kappa_0$ the configurations degenerate into branched
polymers \cite{ambjur}. 

We have interpreted phase A (realized
for large values of $\k_0$) as the CDT remnant of the branched-polymer 
phase, likewise caused by the dominance of the conformal mode.
This suggests that the A-C phase transition may be interpreted as 
a transition where the kinetic term of the 
conformal mode changes sign. This is precisely what happens
at the A-C transition in a Lifshitz diagram, corresponding 
to the expression \rf{2.2} for the free energy of the order parameter $\phi$.  

The effective action for the conformal mode coming out of a 
nonperturbative gravitational path integral receives potential contributions
from several sources: (i) from the bare action (where the kinetic
conformal term has the ``wrong", negative sign), (ii) from the 
measure, and (iii) from integrating out other field components
and, where applicable, other matter fields. It has been argued previously
that the Faddeev-Popov determinants obtained from gauge-fixing the
gravitational path integral contribute 
effectively with the opposite, 
positive sign to the conformal kinetic term \cite{mm,dl}. 
For example, when working in proper-time
gauge, to imitate the time-slicing of CDT, Euclidean metrics can be
decomposed according to\footnote{The conformal decomposition of the
spatial three-metric is essentially unique 
if one requires $g_{ij}$ to have constant scalar curvature.}
\begin{equation}
ds^2=d\tau^2+{\rm e}^{2\phi(\tau,x)} g_{ij}(\tau,x)dx^i dx^j,
\end{equation}
giving rise to a term 
$-1/G^{(b)} {\rm e}^{3\phi}\sqrt{\det g} (\partial_\tau\phi)^2$ 
in the bare gravity Lagrangian density, where $G^{(b)}$ is 
the bare Newton's constant. According to \cite{dl}, 
one expects that the leading contribution from the associated
Faddeev-Popov determinant has the same functional form, but with a
plus instead of a minus sign, and with a different dependence on $G^{(b)}$. 
The presence of contributions of opposite sign 
to the effective action for the conformal mode
$\phi(\tau,x)$ can therefore
lead to two different behaviours, depending on the value of
$G^{(b)}$(equivalently, the $\k_0$ in our model), 
and thus account for the observed
behaviour at the transition 
between phases A and C.\footnote{Related mechanisms 
have earlier been considered in the 
context of purely Euclidean dynamical triangulations by J.\ Smit \cite{js1}.}

In addition to the issues raised above, there is one more question which
we would like to understand in more detail at this stage, which concerns 
the relation of our effective gravitational coupling constant $G$ 
to a more conventional gravitational 
coupling constant, defined directly in terms of coupling 
gravity to matter. It would be desirable to verify that defining
the physical Newton's constant $G$ as the coupling
constant multiplying the effective action for the three-volume, as we have been
doing so far, agrees with a gravitational constant defined more
directly through matter coupling. 
In principle it is easy to couple matter to CDT quantum gravity, as we already know
from multiple studies in the Euclidean case \cite{eumatter}, where spin, scalar and
gauge fields have been considered.
It is less straightforward to come up with a reasonably simple 
set-up for extracting the semiclassical effect of 
gravity on the matter sector (or vice versa),
which is both well-defined on the ensemble of geometries and allows
for effective computer measurements.
Attempts in this direction were already undertaken in the ``old'' Euclidean 
approach \cite{js,newton}, and it is possible that similar ideas can also be used
in our causal version of the theory. As a first step in this direction, the expected
effect of a single point mass coupled to CDT quantum gravity on the
volume profile of the universe has been quantified in \cite{klr}.
Further work on coupled systems of matter and geometry is in progress.

\subsection*{Acknowledgments}
JJ acknowledges partial support by the Polish Ministry of Science  
grant 182/N-QGG/2008/0 ``Quantum Geometry and Quantum Gravity". 
AG has been supported by the Polish Ministry of Science 
grant N~N202~034236 (2009-2010) and N~N202~229137 (2009-2012). 
RL acknowledges support by the Netherlands
Organisation for Scientific Research (NWO) under their VICI
program.

\end{document}